\documentclass{aa}
\usepackage[varg]{txfonts}
\usepackage{afterpage}
\usepackage{capt-of}
\usepackage[caption=false,font=small]{subfig} 
\usepackage{diagbox}

\usepackage{mathtools}
\usepackage[figuresright]{rotating}
\usepackage{hyperref}
\usepackage{amsmath,amssymb}
\usepackage{latexsym}
\usepackage{graphics,graphicx}
\usepackage{epsfig}
\usepackage{multirow}
\usepackage{comment}
\usepackage{appendix}
\usepackage{slashed}
\usepackage{xcolor}
\usepackage{makecell}

\usepackage{booktabs}
\usepackage{tabularx}
\usepackage{wasysym}
\usepackage{color}
\usepackage{enumitem}
\usepackage{cancel}
\usepackage{xspace}

\usepackage[capitalize]{cleveref}

\DeclareRobustCommand{\Sec}[1]{Sec.~\ref{#1}}

\DeclareRobustCommand{\App}[1]{App.~\ref{#1}}

\providecommand{\ie}{\emph{i.e.}\xspace}  
\providecommand{\eg}{\emph{e.g.,}\xspace}

\providecommand{\Gaia}{\emph{Gaia}\xspace}  
\providecommand{\Latte}{\emph{Latte}\xspace}
\providecommand{\insitu}{in situ\xspace}
\providecommand{\FIRE}{\textsc{Fire}\xspace}
\providecommand{\RAVE}{RAVE\xspace}
\providecommand{\mi}{\texttt{m12i}\xspace}
\providecommand{\mf}{\texttt{m12f}\xspace}

\providecommand{\pmra}{\texttt{pmra}}
\providecommand{\pmdec}{\texttt{pmdec}}
\providecommand{\parallax}{\texttt{parallax}}
\providecommand{\radv}{\texttt{radial\_velocity}}
\providecommand{\photg}{\texttt{phot\_g\_mean\_mag}}
\providecommand{\photbp}{\texttt{phot\_bp\_mean\_mag}}
\providecommand{\photrp}{\texttt{phot\_rp\_mean\_mag}}

\providecommand{\LCDM}{$\Lambda$CDM\xspace}

\renewcommand{\vec}[1]{\mathbf{#1}}
\providecommand{\NN}{S}

\providecommand{\Msun}{\,M_{\odot}}
\providecommand{\pc}{\,\text{pc}}
\providecommand{\kpc}{\,\text{kpc}}
\providecommand{\Mpc}{\,\text{Mpc}}
\providecommand{\kms}{\,\text{km/s}}
\providecommand{\cm}{\,\text{cm}}

\providecommand{\FeH}{\text{[Fe/H]}\xspace}

\newcommand {\be} {\begin {equation}}
\newcommand {\ee} {\end {equation}}

\newcommand {\bes} {\begin {equation*}}
\newcommand {\ees} {\end {equation*}}

\usepackage{csvsimple}

\makeatletter
\newcommand\footnoteref[1]{\protected@xdef\@thefnmark{\ref{#1}}\@footnotemark}
\makeatother

\begin{document}

\title{
\large Cataloging Accreted Stars within Gaia DR2 using Deep Learning
}
\author{B. Ostdiek \thanks{Current address: Department of Physics, Harvard University, Cambridge, MA 02138, USA
\email{bostdiek@g.harvard.edu}}
\inst{\ref{inst1}}
\and L. Necib  \inst{\ref{inst2}}
\and T. Cohen \inst{\ref{inst1}}
\and M. Freytsis \inst{\ref{inst3a}}  \inst{\ref{inst3b}}
\and M. Lisanti  \inst{\ref{inst4}}
\and S. Garrison-Kimmmel  \inst{\ref{inst5}}
\and A. Wetzel \inst{\ref{inst6}}
\and R. E. Sanderson \inst{\ref{inst7a}} \inst{\ref{inst7b}}
\and P. F. Hopkins \inst{\ref{inst5}}
}

\institute{
Institute of Theoretical Science, Department of Physics, University of Oregon, Eugene, OR 97403, USA \label{inst1}
\and 
Walter Burke Institute for Theoretical Physics,
California Institute of Technology, Pasadena, CA 91125, USA
\email{lnecib@caltech.edu}
\label{inst2}
\and
Raymond and Beverly Sackler School of Physics and Astronomy, Tel Aviv University, Tel Aviv
69978, Israel \label{inst3a}
\and
School of Natural Sciences, Institute for Advanced Study, Princeton, NJ 08540, USA \label{inst3b}
\and
Department of Physics, Princeton University, Princeton, NJ 08544, USA \label{inst4}
\and
TAPIR, California Institute of Technology, Pasadena, CA 91125, USA \label{inst5}
\and
Department of Physics, University of California, Davis, CA 95616, USA \label{inst6}
\and
Department of Physics and Astronomy, University of Pennsylvania, Philadelphia, PA 19104, USA \label{inst7a}
\and
Center for Computational Astrophysics, Flatiron Institute, New York, NY 10010, USA \label{inst7b}
}

\abstract{}
{The goal of this study is to present the development of a machine learning based approach that utilizes phase space alone to separate the \Gaia DR2 stars into two categories: those accreted onto the Milky Way from those that are \insitu.
Traditional selection methods that have been used to identify accreted stars typically rely on full 3D velocity, metallicity information, or both, which significantly reduces the number of classifiable stars.
The approach advocated here is applicable to a much larger portion of \Gaia DR2.}
{A method known as ``transfer learning" is shown to be effective through extensive testing on a set of mock \Gaia catalogs that are based on the \FIRE cosmological zoom-in hydrodynamic simulations of Milky Way-mass galaxies.
The machine is first trained on simulated data using only 5D kinematics as inputs and is then further trained on a cross-matched \Gaia/\RAVE data set, which improves sensitivity to properties of the real Milky Way.}
{The result is a catalog that identifies $\sim 767,000$ accreted stars within \Gaia DR2.
This catalog can yield empirical insights into the merger history of the Milky Way and could be used to infer properties of the dark matter distribution.}
{}


\titlerunning{Cataloging Accreted Stars using Deep Learning}
\authorrunning{B. Ostdiek, L. Necib et al.}

\maketitle


\section{Introduction} \label{sec:intro}

The two dominant theories of Galactic formation have been monolithic collapse \citep{1962ApJ...136..748E} and a slower build up of merging protogalaxies~\citep{1978ApJ...225..357S}. 
The compatibility of observations with the Lambda--Cold Dark Matter (\LCDM) model, augmented by cosmological simulations, provides strong support for the hierarchical structure formation hypothesis~\citep{1978MNRAS.183..341W}.
The now prevailing view is that a slow and complex process of accretion provides the majority of the dark matter in a galaxy and builds up the stellar halo \citep{1999MNRAS.307..495H,Bullock:2000qf,Bullock:2005pi}.
There is abundant evidence for such accretion events.
For example, the Sagittarius dwarf spheroidal galaxy is currently being disrupted into a large stream that fills the sky \citep{1994Natur.370..194I,2003ApJ...599.1082M}.
The Large and Small Magellanic Clouds \citep{1998MNRAS.299..499W,1999MNRAS.303L...7J}, as well as the Field of Streams \citep{Belokurov:2006ms,2007ApJ...658..337B}, are easily accommodated within this accretion driven framework.

The second release of data from the \Gaia satellite (DR2) \citep{2018arXiv180409365G} allows for the identification of more imprints of accretions, further refining the picture.
It provides parallax and proper motion measurements for over 1.3 billion stars, with high-quality 6D phase-space measurements available for a local subset of over 5 million stars, these multi-dimensional measurements open the door to a deeper understanding of the Milky Way.

Studies of the \Gaia data have already revealed new structures identified as the remnants of the complicated Milky Way accretion history, including \Gaia Enceladus (a.k.a. the \Gaia Sausage) \citep{2018MNRAS.478..611B,2018Natur.563...85H,2018ApJ...863L..28M,2018arXiv180704290L} and Sequoia~\citep{2019arXiv190403185M}, as well as an abundance of new streams~\citep{2018ApJ...860L..11K, 2018MNRAS.475.1537M,2018MNRAS.477.4063M} and disrupted dwarf galaxies \citep{10.1093/mnras/stz1848, 2019arXiv190405370F}.
Unsurprisingly, searches for such accretion events can be highly non-trivial, as they require identifying clusters of accreted stars that share the same origin against the background of the stellar disk.

There have been recent attempts to automate finding structure in the \Gaia data.\footnote{See, \eg \citet{2011MNRAS.415..214M} for related earlier techniques that combine regression techniques with physically-motivated data preprocessing.} 
For instance, assuming a gravitational potential, ~\cite{2018MNRAS.477.4063M} find members of streams contained in 6D flux tubes.
Model independent searches have been done with clustering on the integrals of motion~\citep{2019arXiv190702527B} which also needs the full 6D phase space.
Both of these techniques need to have their halo and accreted stars already identified.
\cite{2019A&A...621A..13V} use a boosted decision tree to classify halo stars, which needs to have metallicity and the full phase space information to do well.
This motivates the use of so-called deep learning techniques, as they are capable of finding non-linear relationships in data using unprocessed low-level inputs \citep[see \eg][for a recent review of deep-learning applications in the physical sciences]{Carleo:2019ptp}.
Applications to the Large Hadron Collider \citep[\eg][]{Larkoski:2017jix}, the Large Synoptic Survey Telescope \citep[\eg][]{2017Natur.548..555H}, many-body quantum systems \citep[\eg][]{2018arXiv180800479Z}, and galaxy morphology \citep[\eg][]{2018ApJ...858..114H, 2019MNRAS.484...93D} make a clear case for the usefulness of deep learning across many disciplines.  
The power of these algorithms is most evident when used on large datasets that contain many non-trivial correlations that can be leveraged to expose interesting structure.

In this paper, we use deep learning to identify accreted stars within \Gaia DR2, which will allow us to boost the size of our sample.
We build a deep neural-network-based classifier, trained on a carefully curated combination of state-of-the-art cosmological simulations augmented by a subset of the \Gaia data itself.
When applied to \Gaia DR2, the result is a high-purity catalog of accreted Milky Way stars, which can be used to search for new structures within our Galaxy.

Of crucial importance to the classifiers developed here are the mock catalogs of \citet{2018arXiv180610564S}.
Because it is possible to identify the accreted population at truth level within these simulated datasets, they allow us to test and optimize our approach before applying the methods to real data.
Additionally, we rely on these mock catalogs to provide the first stage of training, preparing the network for a further stage of training that utilizes actual Milky Way data.
Many efforts in building such mock catalogs already exist.
For example, \texttt{Galaxia} \citep{2011ApJ...730....3S} samples mock stellar halos from \citet{Bullock:2005pi} and mock catalogs based on resampling stars from cosmological simulations have been developed in \citet{1980ApJS...44...73B,1986A&A...157...71R,1987A&A...180...94B,2015MNRAS.446.2274L,2018MNRAS.481.1726G}.
However, with the increase in resolution of hydrodynamic simulations \citep{2014MNRAS.437.1750M,2015MNRAS.454...83W,2016MNRAS.457..844F,2016ApJ...827L..23W, 2017MNRAS.467..179G,Kelley:2018pdy} and the improved modeling of baryonic physics, simulations of the Milky Way can help make self-consistent predictions for survey outputs, especially by utilizing the full knowledge of the formation history.

In this work, we focus on the details of how to use deep neural networks to build a catalog of accreted stars in \Gaia.
The first science results using this catalog are presented in~\citet{catalog_paper}, which uses our catalog to reproduce known structures including \Gaia Enceladus and the Helmi stream and identify potential undiscovered structures.
In particular, we identify a previously-unknown massive stream, Nyx, in the vicinity of the Sun comprising more than 10\% of local accreted stars.
A detailed study of its known properties is presented in~\citet{nyx_paper}.

This paper is detailed and mostly self-contained. 
As such, some readers may prefer to pass over certain sections.
Section~\ref{sec:SimGaiaSky} briefly reviews the \FIRE simulations, with an emphasis on the \Gaia mock catalogs. Our working definition of accreted stars is given in \Sec{sec:accreted_id}.
Section~\ref{Sec:CutAndCount} reviews the traditional selection methods utilized for identifying accreted stars, which will be useful benchmarks to compare to the approach developed here.
Section~\ref{Sec:MLOnFire} provides the machine learning architecture and discusses the training procedure as well as the incorporation of measurement errors.
Section~\ref{sec:Tactics} studies what observables should be used by the neural network.
Section~\ref{sec:GeneralizinginPhaseSpace} finds the optimal observables, assuming all stars are measured well.
Next, \Sec{sec:distance} shows how this changes when trying to generalize to dimmer stars.
Section~\ref{Sec:MLSR} demonstrates that the network can generalize to new view points (as a proxy for new galaxies).
Section~\ref{sec:WhatIsLearned} briefly explores what physical characteristics of the data the machine is using to identify accreted stars.
Section~\ref{Sec:Transfer} introduces the idea of transfer learning, which eventually allows us to train on stars in the Milky Way.
The overall strategy is discussed in \Sec{sec:TransferMethodology}.
In \Sec{sec:TransferExperiments}, we find the optimal empirical labels to use for the transfer process. In addition, we present what network score cut leads to the most consistent results. The strategy and cut are then validated on an independent \FIRE catalog in \Sec{sec:mock}.
Section~\ref{sec:Catalog} describes the application to \Gaia data, producing the  catalog of accreted stars. We first validate the procedure on \RAVE DR5-\Gaia DR2 cross-matched stars in \Sec{sec:CatalogValidation}. The final accreted catalog is presented in \Sec{sec:FinalCatalog}.
Section~\ref{sec:Conc} contains the conclusion.
Appendix~\ref{app:InSitu} shows the stars which the network classifies as \insitu.
Appendix~\ref{app:Errors} provides validation of the error sampling procedure used during the training of the neural network.
Appendix~\ref{app:PhotometricResults} provides an alternative catalog by allowing the network to use photometry as well as kinematics, which is shown in \Sec{sec:distance} not to generalize as well.
We use the Matplotlib~\citep{matplotlib}, Numpy~\citep{numpy}, pandas~\citep{mckinney-proc-scipy-2010}, scikit-learn~\citep{scikit-learn}, and scipy~\citep{scipy} throughout our analysis.

\section{Simulating the Gaia Sky} \label{sec:SimGaiaSky}

Our goal is to develop methods by which we can obtain a catalog of accreted stars.
To do so, we build and test our methodology using mock \Gaia catalogs derived from the \Latte suite of \FIRE-2 cosmological hydrodynamic simulations of Milky Way-mass halos \citep{Wetzel2016,2017arXiv170206148H} --- specifically, those obtained from the simulated galaxies named \mi and \mf from \citet{2018arXiv180610564S}.
This allows us to train and validate our approach on samples where every star's true history is known \citep{Necib:2018igl}.
Below, we briefly review the physics underlying the mocks and describe the algorithm used to identify accreted stars.

\subsection{\FIRE Simulations and the \Gaia Mock Catalogs} \label{sec:fire_mocks}

The numerical methods and physics in the simulations are presented in extensive detail in \citet{2017arXiv170206148H}.
The \FIRE-2 simulations are run with \texttt{GIZMO},\footnote{A public version of \texttt{GIZMO} can be found at \url{http://www.tapir.caltech.edu/~phopkins/Site/GIZMO.html}.} a multi-method gravity and hydrodynamics code \citep{2015MNRAS.450...53H}, using a hybrid tree-PM gravity solver \citep[see \eg][]{2005MNRAS.364.1105S} and the meshless finite-mass (MFM) Lagrangian Godunov solver for hydrodynamics.
Radiative heating/cooling is included over the temperature range $10$--$10^{10}\,\text{K}$, assuming a uniform but redshift-dependent meta-galactic UV background from \cite{2009ApJ...703.1416F}, as well as metal-line, molecular, and other processes.
Star formation occurs only in self-gravitating \citep{Hopkins:2013oba}, Jeans unstable, self-shielding, molecular \citep{Krumholz:2010wm} gas at densities over $1000\cm^{-3}$.
Once stars form, all feedback rates are calculated using standard stellar evolution models \citep[\eg][]{1999ApJS..123....3L}, assuming a \citet{Kroupa:2000iv} initial mass function and the known age, metallicity, and mass of the star particle.
The simulations explicitly model mechanical feedback from supernov\ae{} (Ia and core-collapse) and stellar mass loss (O/B and AGB) \citep{hopkins:sne.methods} as well as multi-wavelength radiative feedback including photoionization and photoelectric heating and radiation pressure from single and multiple scattering \citep{hopkins:radiation.methods}.

Our study relies on two Milky Way-mass galaxies from the \Latte suite of \FIRE-2 simulations, \mi and \mf, introduced in \cite{Wetzel2016,2017MNRAS.471.1709G}.
The simulations adopt a flat \LCDM cosmology with ($\Omega_m$, $\Omega_b$, $h$, $\sigma_8$, $n_s$) = (0.272, 0.0455, 0.702, 0.807, 0.961), with a high-resolution ``zoom-in'' region \citep[\eg][]{onorbe:2013.zoom.methods}, a fully uncontaminated diameter greater than $1.2\Mpc$ at $z = 0$, and a baryonic mass resolution $\approx 7000\Msun$ surrounding the halo of interest in a large cosmological box.
The galaxies \mi and \mf have present-day stellar masses of $5.5$ and $6.9\times 10^{10}\Msun$, respectively, comparable to the Milky Way mass of $(5 \pm 1) \times 10^{10}\Msun$~ \citep{2016ARA&A..54..529B}.
More detailed stellar halo structure comparisons to the Milky Way for both have been previously presented in \citet{2017ApJ...845..101B,2018ApJ...869...12S}.

Mock catalog generation is described in \cite{2018arXiv180610564S}.
Each star particle of mass $\approx 7000\Msun$ is treated as a single-age, single-metallicity population: individual stars are generated by populating the initial mass function according to the particle properties and distributing them within the star particle volume in phase space.
To preserve the wide dynamic range of phase-space densities, the phase-space smoothing kernel is subdivided into 8 age bins for stars formed \insitu, while a separate kernel is used for all accreted stars.
Three different solar positions, each defining a different local standard of rest (LSR), are chosen to construct mocks, spaced uniformly around a circle $R_\odot = 8.2\kpc$ from the galaxy center, defined by the total angular momentum of disk stars.
Lines-of-sight to each star are ray-traced from the solar positions, including extinction and reddening computed self-consistently from the dust and gas in the simulation (and convolved with realistic measurement errors) to compute the observed photometry.

The simulations are not tuned to the Milky Way other than by choosing a dark matter halo of roughly the same mass and choosing from a larger suite of simulations those with similar disk-to-bulge ratios \citep{2018MNRAS.481.4133G} and total stellar masses.
They are, therefore, not perfect Milky Way analogues.
While the total mass (see above) and radial distribution of stars in the disk are quite similar (\eg the radius enclosing 90\% of the stellar mass within $|z| < 1.1\kpc$ is $2.7\kpc$ ($3.4\kpc$) in \mi (\mf), compared to $2.6 \pm 0.5\kpc$ in the Milky Way), the stellar velocity dispersion and scale-height at $R_\odot$ are somewhat larger (\eg the volume-density of stars within $|z| < 200\pc$ for both is $\sim 20 \Msun\pc^{-3}$, versus $\sim 40\Msun\pc^{-3}$ for the Milky Way from \citealt{2016ARA&A..54..529B}).
Likewise, present-day galaxy-wide star formation rates in simulated galaxies are $\sim 3$--$8\Msun\,\text{yr}^{-1}$, compared to $\sim 1.5$--$3\Msun\,\text{yr}^{-1}$ in the Milky Way \citep{2011AJ....142..197C}.

\subsection{Identifying Accreted Simulated Stars} \label{sec:accreted_id}

We derive truth labels for the simulated stars using the method of \citet{Necib:2018igl}.
Our algorithm for tracking the origin of these stars begins by first locating the ones within a Galactocentric distance $r_\text{GC} < 16\kpc$ at the present day.
Stars that do not pass this cut have poor kinematic resolution in  \Gaia DR2, and are therefore not of interest here.
For our purposes, accreted stars are defined as bound to a dark matter  subhalo that fell into the galaxy.
To find such stars, we first define subhalos by applying the \textsc{Rockstar} phase-space finder\footnote{\url{https://bitbucket.org/pbehroozi/rockstar-galaxies}} \citep{2013ApJ...762..109B} to the dark matter  particles alone.
We assign stars to these subhalos by then working backwards from the present day to redshift $z = 99$, spanning 600 individual snapshots of the simulation.
At each snapshot, all stars within the virial radius $R_{200m}$ whose velocity is within $2.5\,\sigma$ of the velocity dispersion for a given subhalo are associated with that subhalo.
If a star is associated with a particular subhalo (other than the main halo) for three out of four consecutive snapshots, it is designated as belonging to that subhalo.

We label these star particles as accreted, and label those that do not pass this selection as \insitu.
To validate this procedure, we also use an alternative approach and apply a cut on the formation distance from the central galaxy at $20\kpc$ ($25\kpc$) on \mi (\mf)~\citep[see][Fig. 1]{2018ApJ...869...12S}.  
This results in a similar categorization into accreted and \insitu stars.
Specifically, for the stars identified by either of these methods, $\sim 74\%~(65\%)$ are selected by both, $\sim 7\%~(17\%)$ are selected only by the high formation distance, and the remaining $\sim 19\%~ (18\%)$ are selected only using the merger history for \mi (\mf).\footnote{The merger history criterion is likely more inclusive due to some stars forming in satellites orbiting within 20--$25\kpc$. 
The distance cut would identify these as \insitu, even though they formed in a different subhalo.}
In the rest of the study, we use the merger history to define whether or not a star has been accreted.

\section{Traditional Selection Criteria}
\label{Sec:CutAndCount}

The primary task of this work is to develop a neural network to distinguish between accreted and \insitu~stars.
To benchmark the success of such networks, we will compare them to standard selection methods used in the literature.
The ``traditional'' approaches take advantage of the fact that the origin of a star is known to be correlated with its stellar position, velocity, and chemical composition.
Relying on simple models for stellar distributions with respect to these inputs, one can motivate a set of selection criteria to identify accreted stars.
In what follows, we benchmark the performance of the  machine against three specific traditional approaches, which we refer to as the \textbf{V} \citep[\eg][]{2010A&A...511L..10N}, \textbf{VM} \citep[\eg][]{2017A&A...598A..58H,2018A&A...615A..70P} and \textbf{ZM selections} \citep[\eg][]{Herzog-Arbeitman:2017fte,Necib:2018igl}.
The methods, along with their names, are explained in more detail below.
Coordinate transformations are performed with Astropy~\cite{astropy:2018}.

\begin{figure*}[t]
  \centering
  \subfloat[Accreted stars.]{
    \includegraphics[trim = {0 0 0 7.1mm},clip,width=0.4\textwidth]{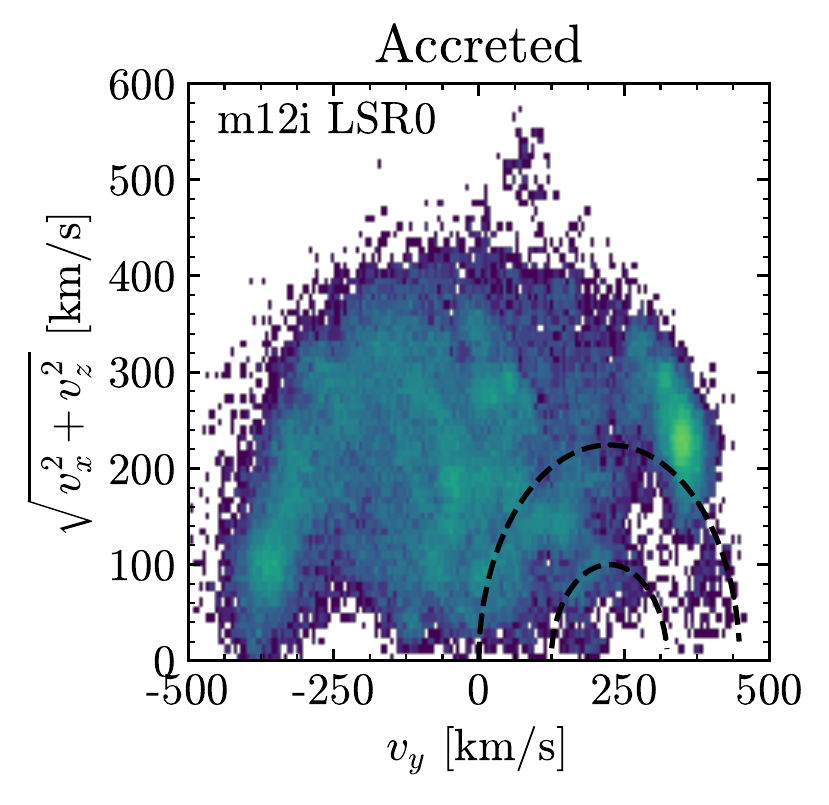}}
  \qquad
  \subfloat[\insitu stars.]{
    \includegraphics[trim = {0 0 0 7.1mm},clip,width=0.4\textwidth]{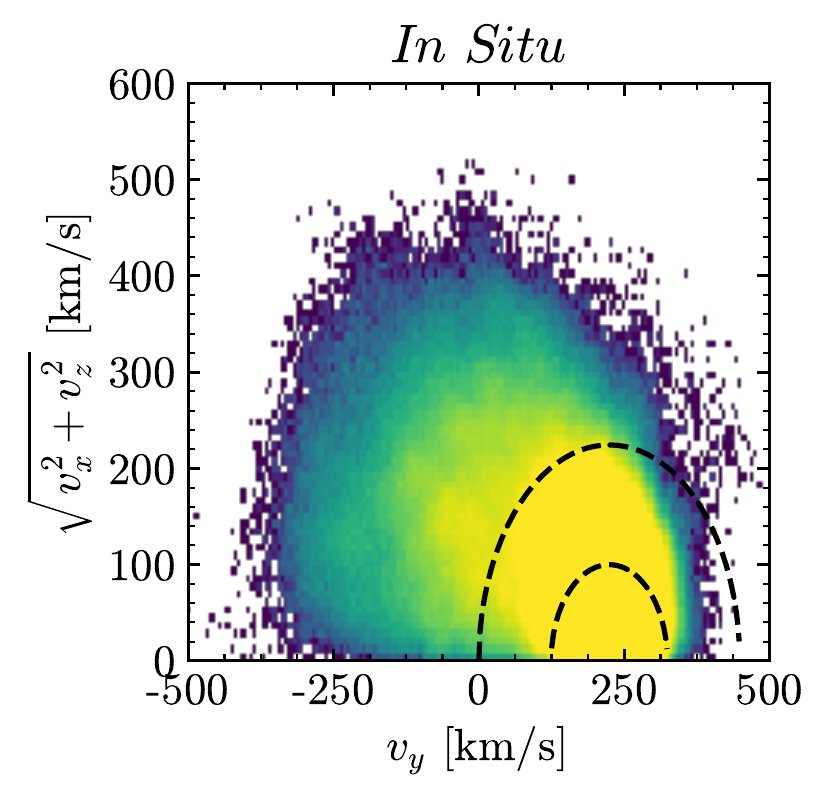}}
    \raisebox{7mm}{\includegraphics[trim = {7.6cm 0 0 0},clip,height=0.325\textwidth]
                                   {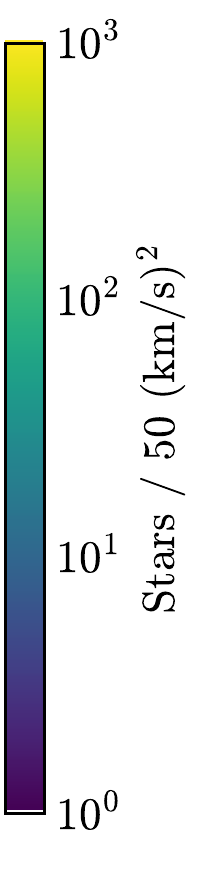}}
  \caption{Truth-level Toomre distributions for LSR0 of \mi with $\delta \varpi /\varpi < 0.10$ and measured $v_r$.  Bright yellow regions denote higher density while the darker blue regions denote lower density, with density ranging from 1--$10^3$ stars/50 (km/s)$^2$.
  The outer (inner) dashed lines denote $|\vec{v} - \vec{v}_\text{LSR}| > 224\kms$ ($124\kms$) with $\vec{v}_\text{LSR} = (0, 224, 0)\kms$.  Velocities are given in Cartesian coordinates centered on the sun, with the $x$ axis oriented away from the Galactic center.
  Accreted stars are essentially uniformly distributed in this plane, although with structure clearly visible, while the \insitu stars peak towards $\textbf{v}_\text{LSR}$, as expected.}
  \label{Fig:ToomreNormal}
\end{figure*}

In \cref{Fig:ToomreNormal}, we plot Toomre diagrams of the truth-level velocity distributions of accreted and \insitu stars for LSR0 of \mi before applying measurement uncertainties.
These plots present the three components of velocity in a Cartesian coordinate system where $v_x$ points from the Galactic center to the sun, $v_y$ points along the direction of the rotation of the Galactic disk, and $v_z$ points towards the angular momentum vector of the disk.
We require that all stars in the mock catalog have a small parallax error ($\delta\varpi/\varpi \le 0.10$) and a measured radial velocity.
We note that for the mock datasets, the parallax error cut also acts an an effective distance cut and removes stars farther than $\sim 4.5\kpc$ away.
The \insitu population is expected to rotate with the disk, consistent with its highest density being near $\vec{v}_\text{LSR} = (0, 224, 0)\kms$.
The accreted population, on the other hand, is relatively uniformly distributed in the Toomre diagram.
It is, however, worth noting the presence of structure in the accreted population, which could be the imprint of particular merger events.

\begin{figure*}[t]
  \centering
  \subfloat[Stars passing \textbf{V selection}]{
    \includegraphics[width=0.3\textwidth]{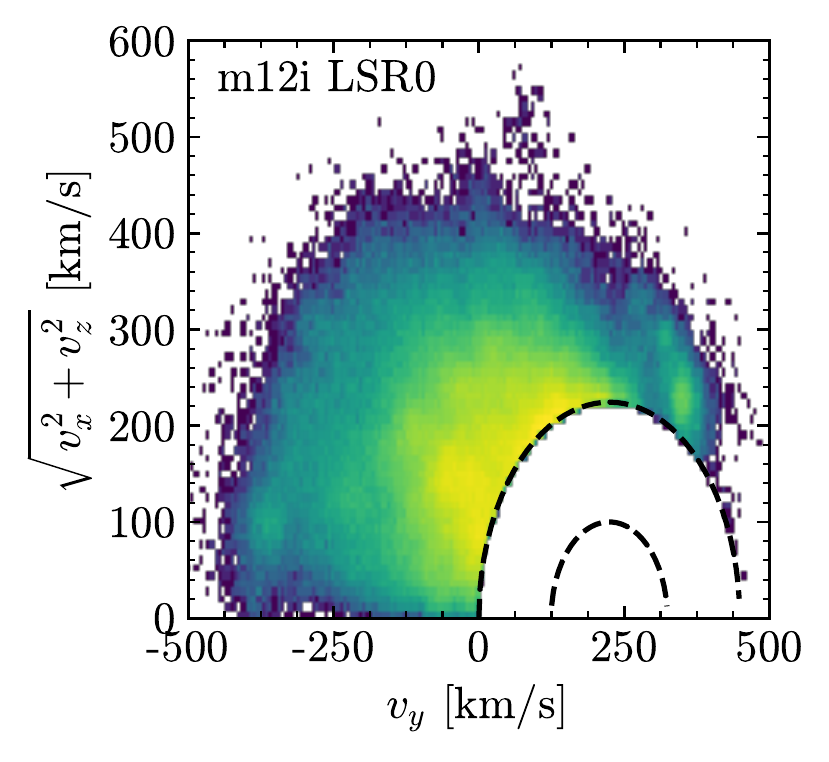}
    \label{Fig:ToomreNormal2V}}
  \hfill
  \subfloat[Stars passing \textbf{VM selection}]{
    \includegraphics[width=0.3\textwidth]{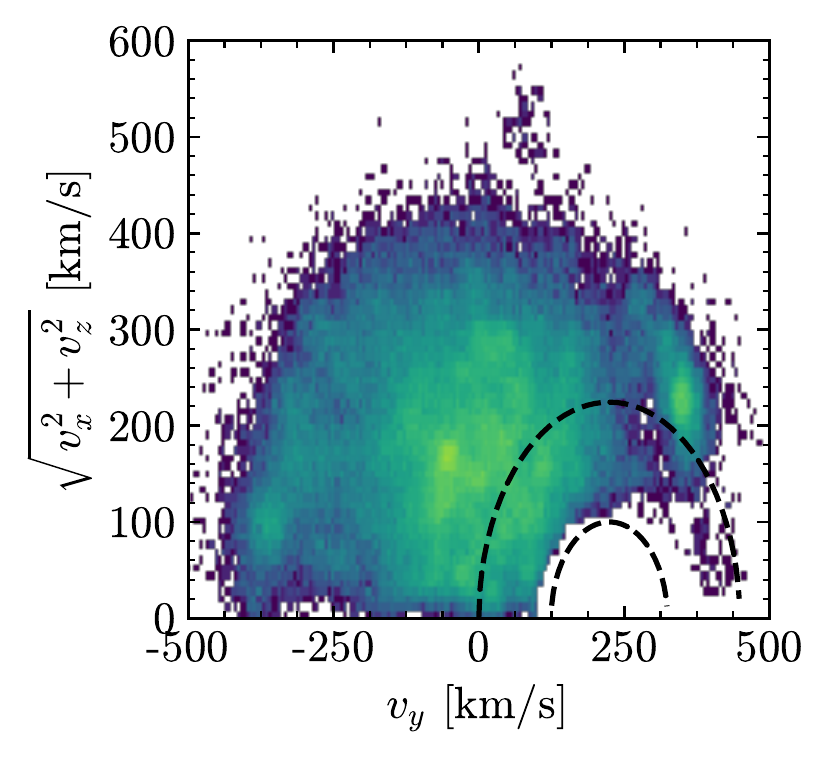}
    \label{Fig:ToomreNormal2VM}}
  \hfill
  \subfloat[Stars passing \textbf{ZM selection}]{
    \includegraphics[width=0.3\textwidth]{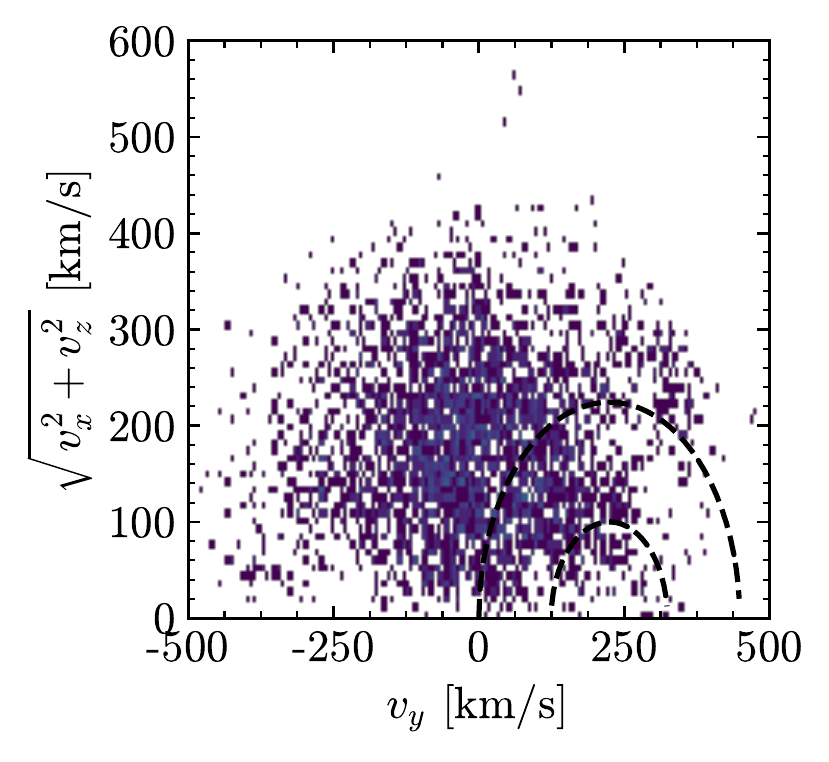}
    \label{Fig:ToomreNormal2ZM}}
  \caption{Toomre diagrams for \mi LSR0 for stars passing traditional cut-based selection criteria, see \cref{Fig:ToomreNormal} for the truth-level plots.  
  The velocity based criterion in \Cref{Fig:ToomreNormal2V,Fig:ToomreNormal2VM} introduces an asymmetry in the Toomre plane, which is absent in \cref{Fig:ToomreNormal2ZM}. 
  Only the \textbf{V selection} is possible with \Gaia DR2 without cross-matching with other catalogs. 
  The purities and efficiencies of all methods are presented in \cref{Tab:TradMethods}.
  }
  \label{Fig:ToomreNormal2}
\end{figure*}

In what follows, we frequently refer to the ``purity" and accreted ``efficiency" of a given selection criterion.
By purity we mean the fraction of stars defined by a given selection criterion that carry an accreted truth label:
\begin{equation}
       \text{Purity} = \frac{N_\text{accreted, selected}}{N_\text{selected}} \,.
\end{equation}
Efficiency is defined as the fraction of all available truth-level stars that are identified, with accreted ($\epsilon_A$) and \insitu efficiency ($\epsilon_I$) defined as: 
\begin{equation}
  \begin{split}
    \epsilon_A &= \frac{N_\text{accreted, selected}}{N_\text{accreted}} \,, \\
    \epsilon_I &= \frac{N_\text{\insitu, selected}}{N_\text{\insitu}} \,.
  \end{split}
  \label{eq:TPR}
\end{equation}
A larger $\epsilon_A$ means that a greater fraction of accreted stars are selected, but one does not want to do this at the expense of picking \insitu stars.
The purity is a measure of this contamination. 

A simple approach to separating the two stellar populations is to apply the \textbf{V selection} criterion \citep{2010A&A...511L..10N}: a star with velocity $\vec{v}$ is defined as accreted if $|\vec{v} - \vec{v}_\text{LSR}| > |\vec{v}_\text{LSR}| = 224\kms$.
\Cref{Fig:ToomreNormal2V} shows the accreted distribution inferred using this criterion.
This method clearly biases the selected accreted population in velocity space as it aggressively removes all stars with velocities close to $\vec{v}_\text{LSR}$, while still presenting significant contamination near the cutoff.
As a result, in this case we find a purity of 9.6\%, but with an accreted efficiency of $\epsilon_A = 86\%$ due to the relatively inclusive selection criterion.
As this method utilizes the 3D velocity, it is only applicable for stars that have line-of-sight velocity measurements in addition to the proper motion measurements.

To improve upon this kinematic-only selection criterion, one can use stellar metallicity as an additional handle.
This is the reasoning underlying the method we refer to as the velocity and metallicity or \textbf{VM selection} \citep{2017A&A...598A..58H,2018A&A...615A..70P}.
This approach is implemented in two stages.
First, the 3D velocity distribution of the stars is fit to a two-component Gaussian mixture model.
One group has a peak consistent with the disk (again $\vec{v}_\text{LSR}$).
The accreted population is defined to be stars having $\FeH < -1$ and more likely to belong to the Gaussian component whose peak velocity is not consistent with the disk.
As is evident from \cref{Fig:ToomreNormal2VM}, this selection criteria allows one to identify accreted stars whose kinematics are more similar to those of the disk.
The accreted efficiency of the resulting selection is again $\epsilon_A = 86\%$ but now with a purity of 30.5\%.
Although the fraction of falsely labeled accreted stars decreases as compared to the \textbf{V selection}, they still make up a majority of the selected sample.
Relative to the \textbf{V selection}, this method can only be applied to a smaller fraction of stars as it requires a metallicity measurement; see \cref{tab:StarNumbers} which provides the relevant statistics for \Gaia DR2.

Both of these approaches demonstrate the intuitive fact that hard cuts on stellar velocities will place a bias on the kinematics of the resulting sample.
Therefore, one must take great care when using the resulting samples as the input to any study that aims to reconstruct the velocity distribution of the accreted stellar population.
An alternative approach is to separate out the \insitu population based on the position and metallicity of the stars alone.
This motivates the $z$ and metallicity or \textbf{ZM selection} \citep{Herzog-Arbeitman:2017fte,Necib:2018igl}, which identifies accreted stars by selecting for those with $|z| > 1.5\kpc$ and $\FeH < -1.5$.
As is clear by inspecting \cref{Fig:ToomreNormal2ZM}, this requirement is the most conservative, with an accreted efficiency of only 2.4\%.
However, for the same mock \Gaia catalog, it leads to a purity of 50.9\% with no obvious bias visible in the Toomre diagram.

\begin{table*}[t]
  \begin{center}
    \renewcommand{\arraystretch}{1.7}
    \setlength{\tabcolsep}{10 pt}
    \setlength{\arrayrulewidth}{.3mm}
    \small
    \begin{tabular}{l|lll|cc}
      Method                & Purity & $\epsilon_A$ & $\epsilon_I$ & $v_{los}$      & \FeH \\
      \hline\hline
      \textbf{V selection}  & 9.6\%  & 86.0\%       & 4.5\%        & \checkmark & \\
      \textbf{VM selection} & 30.5\% & 86.2\%       & 1.1\%        & \checkmark & \checkmark \\
      \textbf{ZM selection} & 50.9\% & 2.4\%        & 0.02\%       &            & \checkmark \\
    \end{tabular}
    \caption{Performance metrics for the traditional selection methods. The middle columns come from the truth-level information of the \mi~LSR0 catalog with $\delta\varpi/\varpi$ $\le 0.10$. The \textbf{V} and \textbf{VM selections} require a measurement of the radial velocity, denoted as $v_{los}$. The purity of a selection is the fraction of selected stars with accreted truth labels. The accreted efficiency ($\epsilon_A$) is the fraction of accreted stars in the sample that get selected and the \insitu efficiency ($\epsilon_A$) is the faction of \insitu stars selected, making $\epsilon_I^{-1}$ the background rejection factor. The right columns show the measurements primarily responsible for limiting available statistics, but which are required to use the method. 
    }
    \label{Tab:TradMethods}
  \end{center}
\end{table*}

The performance of these methods, summarized in \cref{Tab:TradMethods}, is a testament to the challenge of trying to develop unbiased accreted star selection criteria while maintaining high purity and efficiency.
A price is paid in statistics to apply the kinematic selection criteria of these methods.
Specifically, in \Gaia DR2, only 0.3\% of stars have accurate 6D phase-space measurements.
Cross-matching with other observations is required to get metallicities, reducing this to merely 0.02\% of the full dataset in the case of \RAVE DR5, see \cref{tab:StarNumbers} for context.
However, it is worth emphasizing that these issues persist even if full 3D velocity and metallicity measurements are available.
One of the primary benefits of our methodology is its relative flexibility with respect to the completeness of stellar data available.
As we demonstrate below, the machine learning algorithms employed here achieve comparable or better purity of accreted stars, even in cases where the full 6D phase space and/or chemical composition is not available.

\section{Machine Learning on \FIRE} \label{Sec:MLOnFire}

Our goal is to identify accreted stars even in cases where the full phase-space information and/or chemical abundances are unavailable.
To this end, we train a number of dense multilayer feed-forward neural networks as classifiers to distinguish between accreted and \insitu stars. 
Neural networks are chosen over other methods, such as boosted decision trees, for two reasons. 
The first is the general increase in performance. 
The second is the ability to freeze part of the network in transfer learning, which will be discussed in great detail in \Sec{Sec:Transfer}.
%
%
In the bulk of this paper, we will explore the power and limitations of this approach using simulated stellar catalogs, while in \Sec{sec:Catalog} we apply these lessons to data from \Gaia DR2.

\subsection{Neural Network Architecture} \label{sec:NNArch}

Our neural networks are implemented and trained with the \texttt{Keras} package \citep{chollet2015keras} using the \texttt{TensorFlow} backend \citep{tensorflow2015-whitepaper}.
All networks in this study are constructed with five layers (the input, three hidden layers, and the output).
The networks take between 4 and 9 measured quantities per star as inputs --- these variations will be discussed in detail below.
The hidden layers consist of 100 nodes each, using a ReLU activation function, \ie, $h(x) = \max(0,x)$.
The final output layer consists of a single node with sigmoid activation in order to scale the output to lie in the range from 0 to 1.

This network architecture was chosen as the result of a scan on a small set of the simulated data. The width of the layers, and the number of layers was allowed to change. The performance saturated when using three hidden layers of 100 nodes.

The neural network output $\NN(\text{star})$ denotes the value returned when applied to a given star, whose ideal behavior is
\begin{align}
  \NN(\text{star}) =
    \begin{cases*}
      1 & \text{accreted}\\
      0 & \text{\insitu}
    \end{cases*}\,.
\label{eq:labels}
\end{align}
In practice, a continuous range of outputs is produced; larger (smaller) values indicate greater network confidence that a star is accreted (\insitu).
The desired balance of background rejection versus signal efficiency translates into a choice of what value of $\NN(\text{star})$ to use as a selection cut.

Training is performed by adjusting the neural network weights to minimize a loss function, \ie, a function chosen to smoothly vanish when truth labels and network output for a star agree, evaluated on a training set.
Here, we use a weighted version of a standard choice for binary classification, the binary cross entropy:
\begin{equation}
  \mathcal{L} = -\frac{1}{N} \sum_{i=1}^N
                  w_i \Big( y_i \log f_i + \big(1 - y_i\big) \log\big(1-f_i\big) \Big),
  \label{Eq:LossFunction}
\end{equation}
where $y_i$ is the truth label of the star (0 for \insitu, 1 for accreted), $f_i$ is the network prediction, $N$ is the total number of stars in the sample, and $w_i$ is the sample weight of the star (defined below).

\subsection{Training and Error Sampling} \label{sec:TrainingWError}

\begin{table*}[t]
  \begin{center}
    \renewcommand{\arraystretch}{1.7}
    \setlength{\tabcolsep}{4pt}
    \setlength{\arrayrulewidth}{.3mm}
    \small
    \begin{tabular}{lllrr}
      Name                        & Explanation & 
        Unit       & Mean  & $\sigma$ \\
      \hline\hline
      \texttt{l}                  & Galactic longitude & 
        [deg]      & 2.72  & 78.2     \\
      \texttt{b}                  & Galactic latitude & 
        [deg]      & 0.71  & 26.1     \\
      \texttt{pmra}               & Proper motion in right ascension &
        [mas/year] & -1.66 & 10.6     \\
      \texttt{pmdec}              & Proper motion in declination &
        [mas/year] & -2.83 & 11.0     \\
      \hline
      \texttt{parallax}           & Parallax &
        [mas]      & 0.48  & 0.86     \\
      \hline
      \texttt{phot\_g\_mean\_mag}  & Extincted apparent $G$-band mean magnitude & 
        [mag]      & 18.49 & 2.0      \\
      \texttt{phot\_bp\_mean\_mag} & Extincted apparent $G_{B_p}$-band mean magnitude & 
        [mag]      & 19.16 & 2.3      \\
      \texttt{phot\_rp\_mean\_mag} & Extincted apparent $G_{R_p}$-band mean magnitude & 
        [mag]      & 17.76 & 1.9      \\
      \hline
      \texttt{radial\_velocity}    & Line-of-sight velocity & 
        [km/s]     & -6.15 & 75.5     \\
      \texttt{feh}                 & [Fe/H] &
        [dex]      & -0.20 & 0.42  \\
    \end{tabular}
    \caption{Inputs used for star classifications. The mean and standard deviation come from the \mi LSR0 mock catalog derived from the \FIRE simulation and are used to rescale the measurements such that for \mi LSR0 all network inputs have a mean of zero and unit variance. These scale factors are applied to all the datasets used in this work: \mi, \mf, and \Gaia DR2.}
    \label{Tab:Normalizations}
  \end{center}
\end{table*}

To develop a robust classifier, we incorporate all uncertainties in the input variables into our training methodology.
This avoids letting the classifier learn to heavily rely on kinematic or photometric properties of stars that are not actually well-measured.
Additional justification for our error sampling approach is provided in \App{app:Errors}.

Because the \insitu stars outnumber the accreted stars by a factor of $\sim 100$ for stars with  $\delta\varpi/\varpi < 0.10$, we introduce compensating weights for the two populations.\footnote{If the full catalog were used, the ratio would be smaller, especially at larger distances.}
A larger weight for accreted stars ensures that the network is more sensitive to their properties during training. 
However, too large a weight would risk making the algorithms too sensitive to statistical fluctuations in the accreted subsample. 
Balancing between these concerns, we introduce a compensating factor of 5 in the weights relative to their proportion in the training set,
\begin{equation}
  \begin{split}
    w_\text{accreted} &= \frac{1}{5} \frac{N_\text{total}}{N_\text{accreted}}\,, \\[7pt]
    w_\text{\insitu} &= \frac{N_\text{total}}{N_\text{\insitu}}\,.
  \end{split}
\end{equation}
In addition, we set the size of the training batches so that there is an average of 5 accreted stars in each batch.
This leads to a batch size of 915 for data sets using $v_{los}$ measurements, and one of 565 when $v_{los}$ is not used.
We do not perform a comprehensive optimization over these choices, since we confirmed that the results are not very sensitive to the specific choice of relative weights or batch size.
However, we note that the \texttt{Keras} default batch size of 32 yielded very inefficient training.

The training proceeds as follows.
A given mock catalog is split into three subsets: one for training, one for validation, and one for testing.
The validation and testing subsets each have a fixed size of 1 million unique stars.
The training set is comprised of the remaining stars, in all cases consisting of at least 9.3 million stars.

For each star evaluated during training, we incorporate the impact of finite errors by generating 20 new instances of that star from a normal distribution whose center is the observed value and whose standard deviations are taken to be the observational uncertainties.
The input variables for the stars (including all the ``stars'' generated by the random sampling procedure) are then rescaled so that the inputs are all roughly the same size.
Specifically, the rescaling is chosen such that (over the stars of \mi LSR0 mock catalog with $\delta \varpi /\varpi < 0.10$) each input has a mean of 0 and a variance of 1.
\Cref{Tab:Normalizations} provides the full list of inputs, as well as the mean and variance used for the rescaling.
Note that this rescaling is performed with the values in \cref{Tab:Normalizations} for all the datasets used in this work: \mi, \mf, and \Gaia DR2.  Before training, the network weights are initialized with the default \texttt{glorot\_uniform} method~\citep{glorot2010understanding}, which ensures that the initial variance of the inputs to each node is independent of its position in the network.

An epoch of training consists of two parts.
First, we perform one iteration over all the batches in the training set, decreasing the loss function given in \cref{Eq:LossFunction} for each batch in turn by updating the weights using the \texttt{Adam} optimizer~\citep{DBLP:journals/corr/KingmaB14}. The initial learning rate is set to $10^{-3}$.
The learning rate controls how much the weights change.
Then, after iterating over all of the batches in the training set, a similar procedure is performed using the validation set, including the random sampling over the observational errors.
However, only the loss is calculated --- the network weights are not updated during validation.
This defines one epoch of training.
The stars in the training set are then randomly shuffled and the next epoch begins.

If the loss calculated on the validation set does not improve for 5 epochs of training, the learning rate is reduced by a factor of 10, but is not allowed to decrease below $10^{-6}$.
Training ends when the validation loss has not improved for 10 epochs.  
This procedure typically takes $\sim 50$ epochs to complete.
Specifying training completion in terms of validation set loss allows us to ensure that the network  has the best potential to generalize to stars that it was not trained on. 
The value of the loss between the validation and training sets was comparable so we did not find it necessary to implement additional procedures that help avoid overfitting such as regularization or dropout.
Once the network weights are fixed, all performance metrics are computed on the test set.

\section{Tactics for Hunting Accreted Stars} \label{sec:Tactics}

In this section, we explain our approach for optimizing the neural network configuration.
Through extensive testing on the mock catalogs, we address the question of how many input measurements (and what error tolerances) are required to effectively yield a large-statistic, high-purity sample of accreted stars.
First, we will explore the impact of eliminating the network's access to one or more of the six phase-space degrees of freedom (d.o.f.).
An optimal selection of inputs is critical because \Gaia does not measure all six d.o.f.\ with the same level of accuracy.
Then, we will investigate to what extent including photometry and/or metallicity into the analysis improves the network's ability to discriminate.
Finally, we will discuss how well the network generalizes from one viewpoint of a simulated galaxy to another.  
This informs how well we can expect a network trained on simulation to behave when applied to the actual \Gaia dataset.
Further aspects of generalization to other simulations and \Gaia data itself are discussed in \Sec{Sec:Transfer}.

\subsection{Performance Versus Input Dimensionality} \label{sec:GeneralizinginPhaseSpace}

We begin by considering how performance is affected when greater or fewer input phase-space d.o.f.\ are provided to the network.
We use three choices of kinematic networks:
\begin{itemize}[leftmargin=10pt]
  \item \textbf{4D}: \texttt{l}, \texttt{b}, \pmra, \pmdec
  \item \textbf{5D}: \texttt{l}, \texttt{b}, \pmra, \pmdec, \parallax
  \item \textbf{6D}: \texttt{l}, \texttt{b}, \pmra, \pmdec, \parallax, \radv
\end{itemize}
In addition, we also test the impact of providing the networks with access to photometry or metallicity:
\begin{itemize}[leftmargin=10pt]
  \item \textbf{Kinematic + Photometric}: \texttt{l}, \texttt{b}, \pmra, \pmdec, (\parallax, [\radv]), \photg, \photbp, \photrp
  \item \textbf{Kinematic + [Fe/H]}: \texttt{l}, \texttt{b}, \pmra, \pmdec,\\ (\parallax, [\radv]), \texttt{feh}
\end{itemize}
where by (\parallax, [\radv]) we denote that the kinematics will be specified as 4D, 5D, or 6D (as defined above) in the results.

\begin{figure*}[t]
  \centering
  \includegraphics[width=0.95\textwidth]{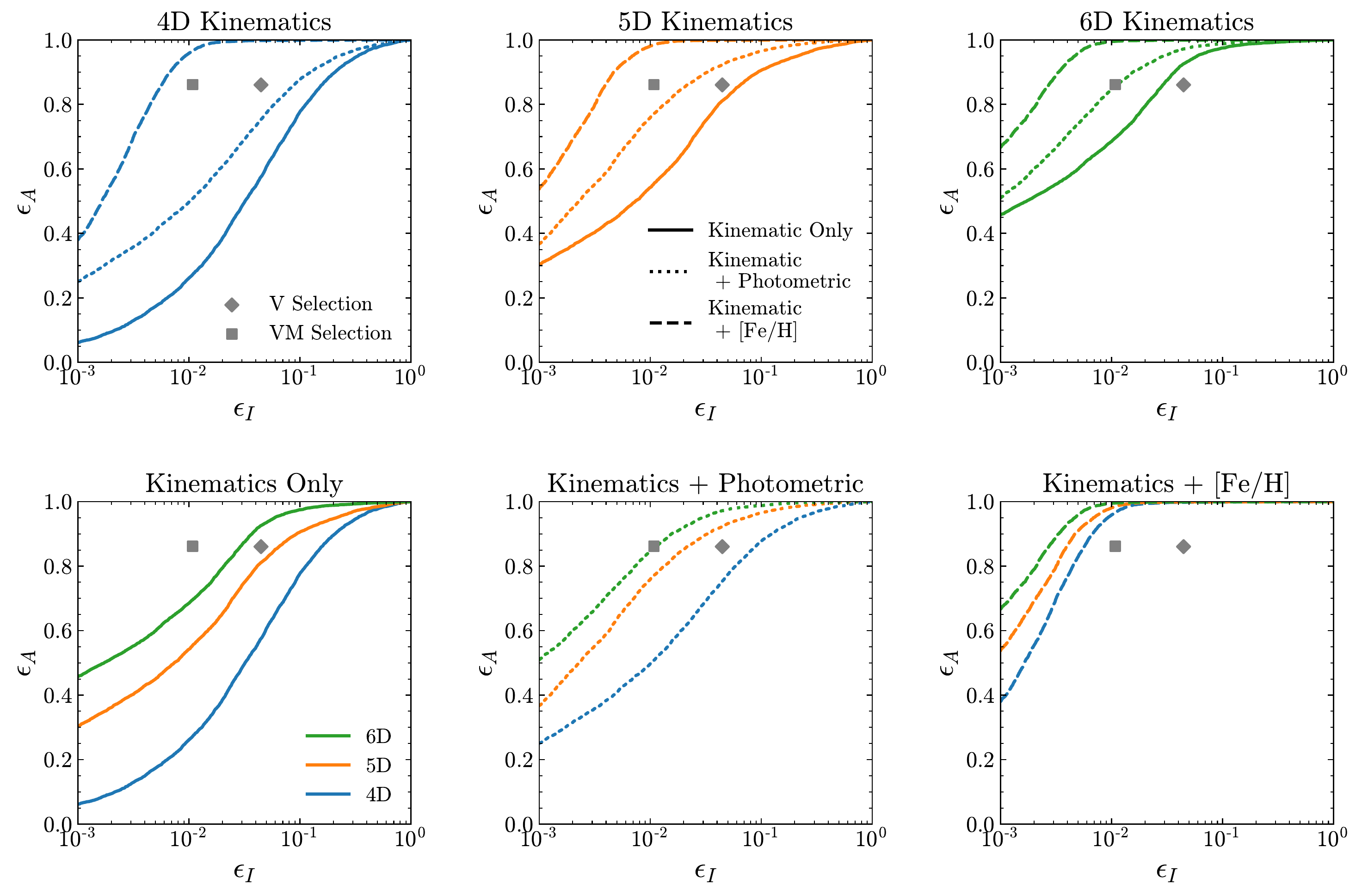}
  \caption{Metrics on the test set of stars from the \mi LSR0 \FIRE mock catalog with $v_{los}$ measurements and $\delta \varpi/\varpi < 0.10$.
  Such ROC curves compare the accreted and \insitu efficiencies, as defined in \cref{eq:TPR}, for different networks.
  In general, a network has better performance if the accreted efficiency is high, and the \insitu efficiency is low, \ie, if the ROC curve tends to the top left of a given plot. 
  In this figure, the label on each curve denotes what information the network had access to at training, and the plot title indicates the common network input for all curves in that plot.
  Note that the same nine curves are plotted in the top row and the bottom row --- they are only organized differently to make the comparisons as transparent as possible.
  The gray symbols denote the performance for the \textbf{V} and \textbf{VM selections} described in \Sec{Sec:CutAndCount}.
  \textbf{ZM} is out of the range of the plots, as it has an \insitu efficiency of $1.8\times10^{-4}$ and an accreted efficiency of $3.3\times10^{-2}$.
  The network performance consistently beats the traditional methods in all cases where metallicity is included as an input, even if only 4D kinematics are available.
  The addition of photometric inputs provides a relative improvement compared to kinematics-only networks, although it is not as powerful as metallicity.
  }
  \label{Fig:SmallSetNetCompare}
\end{figure*}

To compare networks, we use the receiver operating characteristic (ROC) curve.
A ROC curve is a parametric curve generated by comparing the accreted (signal) versus \insitu (background) efficiency as the cut on the network output, which measures the network's confidence in its classification, is scanned from 0 to 1.
Note that for a given background rejection factor ($\epsilon_I^{-1}$), a larger accreted efficiency $\epsilon_A$ represents a better classifier, so that an ideal classifier has a ROC curve that asymptotes to the top left corner of the plots, \eg see \cref{Fig:SmallSetNetCompare}.
Multiple
summary statistics for binary classifiers are available including area
under the curve, $F_1$ score, Matthews correlation coefficient and
others.
Especially in imbalanced classification problems such as this,
relying on a single metric combines differing sources of
misclassification that can require markedly different treatments in
subsequent physics analyses.
We therefore present full ROC curves
almost exclusively throughout this paper.

We compare the ROCs for various networks to the performance of the traditional selection methods introduced in \Sec{Sec:CutAndCount}.
In this way, we can assess what input variables are needed to match or exceed their performance.
As these traditional methods define a unique signal region, they do not yield a continuum of operating points.
Therefore, each is represented as a single point on a ROC plot.
Due to the requirements of most of these traditional methods, we are only able to apply them to a subset of the mock stars from the \mi LSR0 catalog: those with parallax measurement errors $\delta\varpi/\varpi < 0.10$ and radial velocities $v_{los}$.  
Therefore, all of the results (both machine learning and cut-based) in this sub-section are performed using only this 6D subset.

\begin{table}[t]
  \begin{center}
    \renewcommand{\arraystretch}{1.7}
    \setlength{\tabcolsep}{4pt}
    \setlength{\arrayrulewidth}{.3mm}
    \small
    \begin{tabular}{l|ccc}
      $\delta\varpi/\varpi < 0.10$; $v_{los}$
                     & $N_\text{accreted}$ & $N_\text{\insitu}$ & Purity \\
      \hline\hline
      Training set   & 51,209              & 9,349,637                 & 0.55\% \\
      Validation set & 5,364               & 994,636                   & 0.54\% \\
      Test set       & 5,462               & 994,538                   & 0.55\% \\
    \end{tabular}
    \caption{Detailed composition of stars within the \mi LSR0 dataset with a measurement of radial velocity $v_{los}$ and parallax error $\delta\varpi/\varpi < 0.10$.
    The stars are divided into training, validation, and testing sets when utilized for training the neural network and testing its subsequent performance.
    For each set, we provide the number of true accreted and \insitu stars as well as the purity, which is defined as the number of truly accreted stars selected divided by the total number of  stars in the particular dataset.}
    \label{Tab:SmallSet}
  \end{center}
\end{table}

\Cref{Tab:SmallSet} gives the division of the mock data into training, validation, and test sets.
We train a series of neural networks following the procedure of \Sec{sec:TrainingWError}, so that performance using the validation set are what determines when training is completed.
Then ROC curves are computed using the test set.
The resulting behavior is shown in \cref{Fig:SmallSetNetCompare} for the different networks.
The upper panels show the results for networks that see 4D (on-the-sky position and velocity), 5D (adding parallax), or 6D (adding radial velocity) kinematics, from left to right.
Within each panel, the solid, dotted, and dashed lines represent networks that use only kinematics, kinematics + photometry, or kinematics + metallicity, respectively.
The grey markers display the working points for the cut-based methods.
(The \textbf{ZM selection} lies outside the plot range, but is given in \cref{Tab:TradMethods}.)

Using 4D kinematics alone (\cref{Fig:SmallSetNetCompare}, upper left panel), the neural network does not achieve the level of classification obtained by the traditional methods.
However, by adding in the photometric inputs, the performance dramatically improves because the  photometric magnitude is correlated with the distance (or parallax) and the metallicity.  
Furthermore, since most of the accreted stars in \mi have lower metallicity than the \insitu stars, having direct access to \FeH yields even better network performance.

Next, we train a network to use 5D kinematic inputs, where the parallax measurement is now made available.
The bottom left panel of \cref{Fig:SmallSetNetCompare} shows that the 5D kinematics-only network using parallax has significantly better performance than the 4D kinematics-only case.
As shown in the upper middle panel of \cref{Fig:SmallSetNetCompare}, including parallax in the kinematics-only network allows for performance close to that of the \textbf{V selection}. 
When the additional photometric information is added, the network can nearly achieve the performance of the \textbf{VM selection}.
As parallax and photometric information is available for a large fraction of the stars in \Gaia DR2, this initially appears to be a very encouraging result.
However, as we demonstrate below, networks that utilize photometric information do not generalize as well, and so the primary focus of this paper will be on the kinematic networks.  

Finally, the right panel in the upper row of \cref{Fig:SmallSetNetCompare} includes the radial velocity information, giving the network access to the complete 6D phase space.\footnote{It would also be interesting to examine other combinations of input variables. For example, as parallax becomes harder to measure beyond $\sim 4\kpc$, a 5D combination of \texttt{l}, \texttt{b}, \pmra, \pmdec, and \radv{}  could allow for access to stars that are farther away. Going to a 3D model with \texttt{l}, \texttt{b}, and \radv{} could extend the reach in distance even further. However, the method chosen here is applicable to a larger number of stars, so we leave an exploration of these other choices for future studies aimed at subsequent \Gaia releases and spectral surveys.}
Unsurprisingly, having 6D information allows the networks to achieve excellent distinguishing power between accreted and \insitu stars.
When combined with photometric or metallicity information, the resultant networks outperform the standard methods.
In all cases the networks provide superior performance when given access to equivalent information.
Unfortunately, line-of-sight velocity measurements are currently only available for a small fraction of stars in \Gaia DR2, see \cref{tab:StarNumbers}.
However, these results certainly motivate updating our analysis for future \Gaia data releases (as well as ground-based spectroscopic surveys such as LAMOST~\citep{2012RAA....12..735D}, DESI~\citep{Aghamousa:2016zmz}, and SDSS-V~\citep{2017arXiv171103234K} that will include more stars with 6D measurements.

The bottom row of panels in \cref{Fig:SmallSetNetCompare} contain the same nine curves, but organized by the extra information provided to the network: kinematics only, kinematics + photometric, and kinematics + metallicity, respectively.
The different curves now indicate the cases of 4D, 5D or 6D kinematics.
While adding more information is clearly better, this makes it clear that systematically there is a larger gain in going from 4D to 5D than from 5D to 6D, especially when only the kinematic information is provided (left bottom panel).

These results demonstrate that deep learning can exploit hidden correlations in the data to identify accreted stars.
That being said, it is not obvious how this approach can generalize if the test/validation sets include incomplete information about the stars, or when applying the networks trained on one location to another viewpoint within a simulated galaxy.
We turn to addressing these questions next.

\subsection{Generalizing from Brighter to Dimmer Stars} \label{sec:distance}

\begin{table}[t]
  \begin{center}
    \renewcommand{\arraystretch}{1.7}
	\setlength{\tabcolsep}{4pt}
	\setlength{\arrayrulewidth}{.3mm}
	\small
    \begin{tabular}{l|ccc}
      $\delta \varpi/\varpi < 0.10$
                     & Accreted & \insitu &  Purity \\
      \hline\hline
      Training set   & 430,376  & 48,266,382     & 0.88\%  \\
      Validation set & 8,938    & 991,062        & 0.89\%  \\
      Test set       & 8,828    & 991,172        & 0.88\%  \\
    \end{tabular}
    \caption{Same as \cref{Tab:SmallSet}, but with no requirement of $v_{los}$, \ie, requiring parallax errors $\delta\varpi/\varpi < 0.10$ only.
    Removing the requirement on $v_{los}$ increases the sample size by a factor of $\sim 5$ and the number of accreted stars by a factor of $\sim 9$.}
    \label{Tab:MediumSet}
  \end{center}
\end{table}

The previous subsection focused on the subset of stars with small parallax errors and radial velocities.
We now consider how the network performance is affected if we train a network on the subset of stars that are closest to the viewpoint (and hence satisfy the requirement of having $\delta\varpi/\varpi < 0.10$ and a measurement of $v_{los}$) using only 5D kinematic information, and then apply the network to a larger data set consisting of stars that are farther away and therefore are not measured as well.
Comparing the star counts in \cref{Tab:SmallSet} and \cref{Tab:MediumSet} reveals that the size of the mock dataset increases by about a factor of 5 by relaxing these quality cuts.
However, this increase is not evenly distributed between accreted and \insitu stars; the fraction of truth-level accreted stars within the training set increases from $0.55\%$ to $0.88\%$.
We will see that indeed the network is learning general enough features for this extrapolation to yield useful results, which will ultimately bolster our ability to produce a high statistics sample of accreted stars.

In \Sec{sec:GeneralizinginPhaseSpace}, we demonstrated the extent to which using 5D rather than 4D kinematics improves network classification performance.  
Repeating this test on the mock data with no $v_{los}$ requirement, the conclusions are unchanged.
Therefore, we only show the results for 5D kinematics going forward.

\begin{figure}[t]
    \centering
    \includegraphics[width=0.4\textwidth]{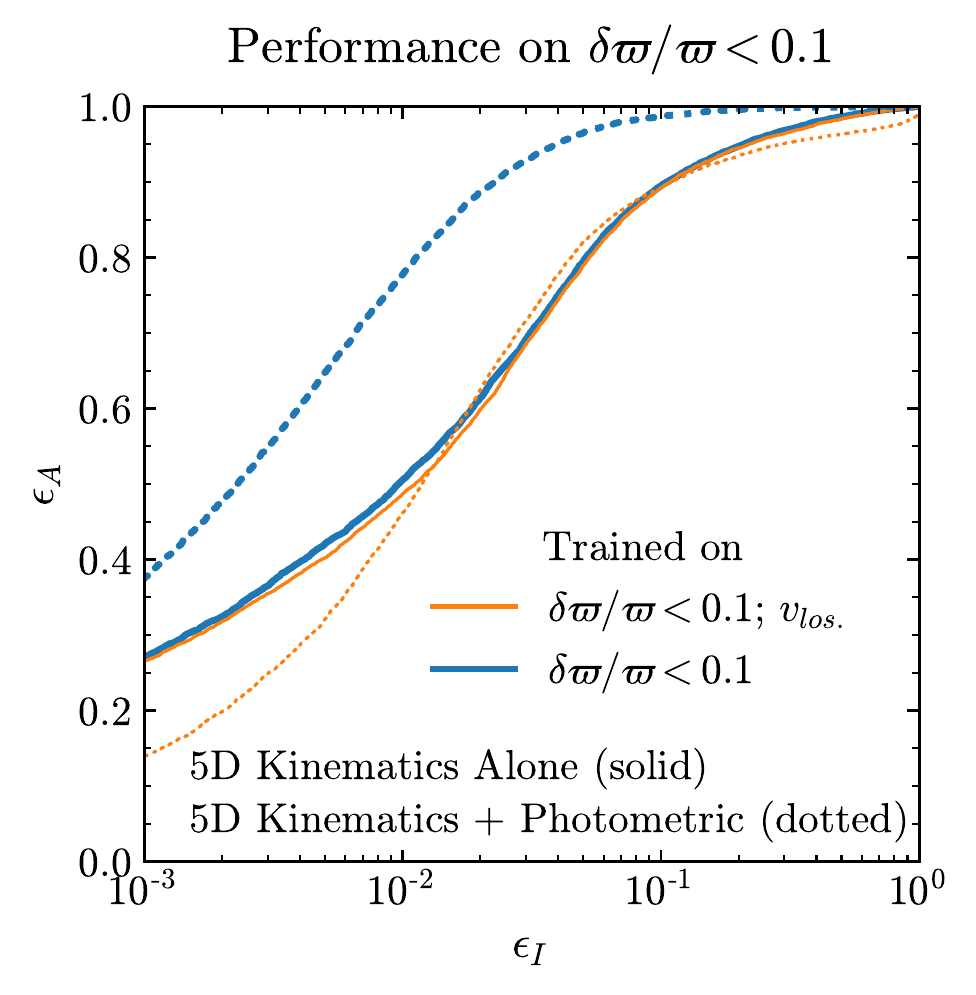}
    \caption{Testing how networks trained on nearby or bright stars generalized to farther or dimmer stars, using test data of the \mi LSR0 \FIRE mock catalog.
    We compare results where the network is trained on stars with $\delta \varpi/\varpi < 0.1$ with either $v_{los}$ measurements required (orange) or not (blue).
    Solid or dotted lines indicate if the network uses 5D kinematics or also includes photometry as inputs.
    All networks are tested on the 5D dataset, such that stars have a small parallax error, but they may or may not have a radial velocity measurement.  
    The network that only uses 5D kinematics gives equivalent results regardless of whether it is trained on data with radial velocities or not, \ie, the solid blue and orange lines are comparable.
    However, we find that when 5D kinematics + photometry are used as inputs, the network performance is significantly hampered when training on the data set with radial velocities and testing on the broader data set with no $v_{los}$ requirement, \ie, the dotted orange line is suppressed relative to the dotted blue line.
    }
    \label{Fig:MediumSetNetCompare}
\end{figure}

The ROC curves for the 5D networks are presented in \cref{Fig:MediumSetNetCompare}.
The blue lines correspond to the networks that were trained on stars with $\delta \varpi/\varpi < 0.10$.
In contrast, the networks depicted by the orange lines were trained on stars that were additionally required to have a line-of-sight velocity measurement, \ie, training only involves the more nearby stars.
Note that these results are different than those presented in \cref{Fig:SmallSetNetCompare}, because the network is now applied to test stars which may or may not have line-of-sight velocities.

\begin{figure}[t]
    \centering
    \includegraphics[width=0.4\textwidth]{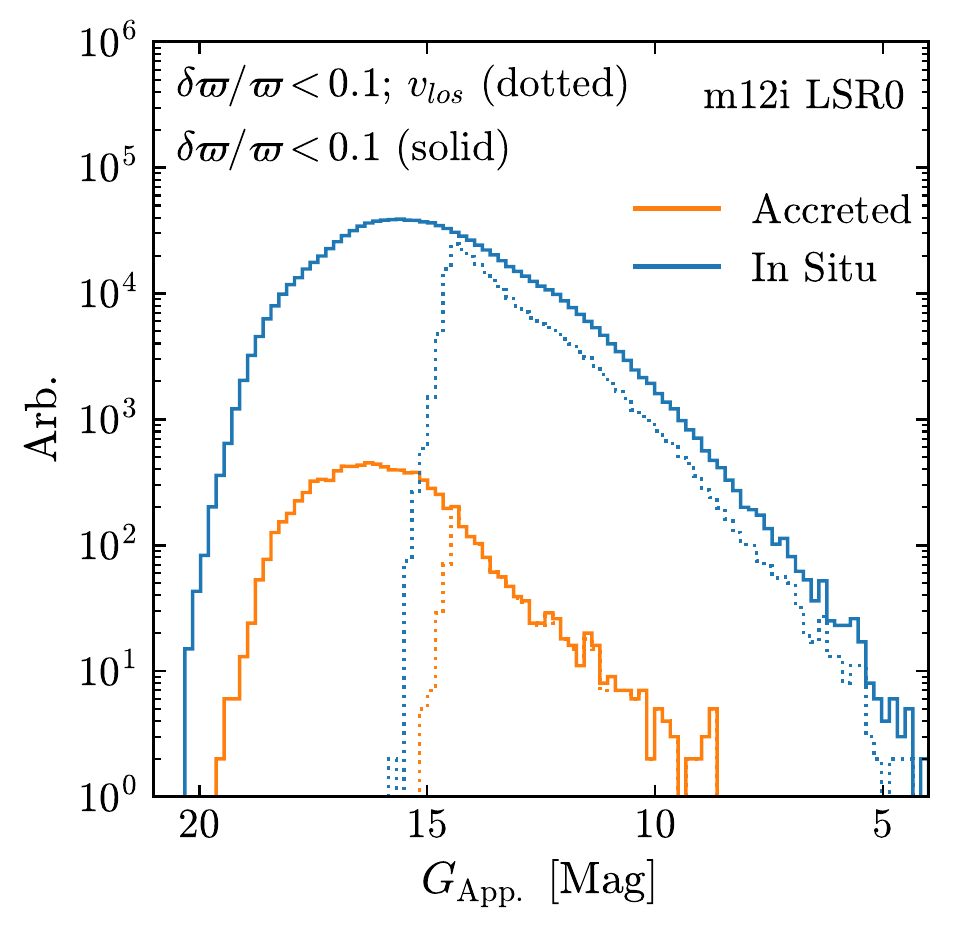}
    \caption{Distributions of the mean apparent brightnesses in the $G$ band for \cref{Tab:SmallSet,Tab:MediumSet}. 
    The normalization of the distribution is in arbitrary units.
    In situ stars are indicated by the blue lines, and accreted stars are shown as orange lines.
    Networks that are trained on stars with $\delta \varpi/\varpi < 0.10$ and $v_{los}$ measurements (dotted lines), using both 5D kinematics and photometry as inputs, do not perform well when tested on all stars with $\delta \varpi/\varpi < 0.10$ (solid lines).
    This result was demonstrated in \cref{Fig:MediumSetNetCompare}, and here we expose the root of the problem.
    A network that is trained on stars where $v_{los}$ is required only sees relatively bright stars and does not generalize well when applied to a data set where dimmer stars are present.    
    }
    \label{Fig:Photometric}
\end{figure}

First, we highlight the two solid lines in \cref{Fig:MediumSetNetCompare}, which only use 5D kinematic information as network inputs.  
The orange and blue solid lines are essentially indistinguishable.
This indicates that there is little drop in performance when the requirement of a radial velocity measurement is imposed, which makes sense when the network is only trained using 5D kinematics.
Looking forward, this makes it plausible that we can use the small subset of stars within \Gaia DR2 that have radial velocity measurements to train a network on actual stars.
If the network only uses the 5D kinematic information for the stars, the network will safely generalize to stars with no radial velocity measurements.

Next, we consider what happens when photometry is included in the training (dotted lines in \cref{Fig:MediumSetNetCompare}).
When training and testing on stars that may not have radial velocities (blue), the inclusion of photometric data improves the performance of the network, as anticipated from the previous section.
However, something surprising happens when the network is trained on stars with radial velocity measurements and is then applied to the less restrictive data set (orange line).
Contrary to expectations, this network performs the worst when high purity is demanded.
The explanation for this can be inferred from the apparent brightness plots for the two mock data sets in \cref{Fig:Photometric}.
This figure shows that only bright stars have a line-of-sight velocity measurement.
Since the network that is trained on a more restrictive dataset and does not see stars with an apparent magnitude above 15, it does not properly identify the dimmer stars in the expanded data set.

Additionally, we also computed the ROC curves for stars in the test set with $\delta\varpi / \varpi < 0.1$ and $G_{\rm{App.}} < 15$ but no restriction on $v_{los}$.
There are roughly twice as many stars is this set than when requiring $v_{los}$, and the results generalize better between training on stars with or without $v_{los}$ requirements. 
However, this only applies to less than half of the stars with $\delta\varpi / \varpi < 0.1$.

We will eventually use transfer learning to train the network on stars in the Milky Way using empirical labels which require spectroscopy. 
This motivates us to focus our attention on the 5D kinematic network.
Going forward, the networks will not be given access to photometric information.
Appendix~\ref{app:PhotometricResults} shows the resulting analysis when using the network with photometry.

The results of this section have demonstrated that it is possible to use the measurements available for a large portion of \Gaia DR2 to separate \insitu and accreted stars.
Next, we will explore the extent to which a network trained on one location within a galaxy can be applied to another.

\subsection{Exploring Multiple LSRs}
\label{Sec:MLSR}

Three synthetic \Gaia surveys are provided for \mi \citep{2018arXiv180610564S}, each separated in the plane of the disk by angular intervals of $2\pi/3$.
These surveys allow one to study the effects of features like the bar and spiral arms, which break rotational symmetry.
Moving between these viewpoints provides a way to validate the level to which the networks rely on features that are specific to a particular LSR.
One might be concerned that the neural network is performing so well by memorizing localized substructures.
Since moving between LSRs causes these substructures to occupy very different regions of the 5D phase space (and possibly removes them), passing this test tells us that our algorithms are learning gross features of the galaxy, as opposed to merely remembering the particulars of a specific viewpoint.

Thus far, all results have utilized mock data taken from \mi LSR0.
Now we will assess the impact of moving to a different position using the surveys of \mi denoted by LSR1 and LSR2.
We again select stars with 5D kinematics (small parallax errors and no requirement on line-of-sight velocity measurements).
From this subset, we then assign 1~million stars for validation and 1~million for testing.
For reference, we note that the LSR1 and LSR2 datasets are slightly larger than that of LSR0; the stars in each survey are unique.

\begin{figure*}[t]
  \centering
  \includegraphics[width=0.95\textwidth]{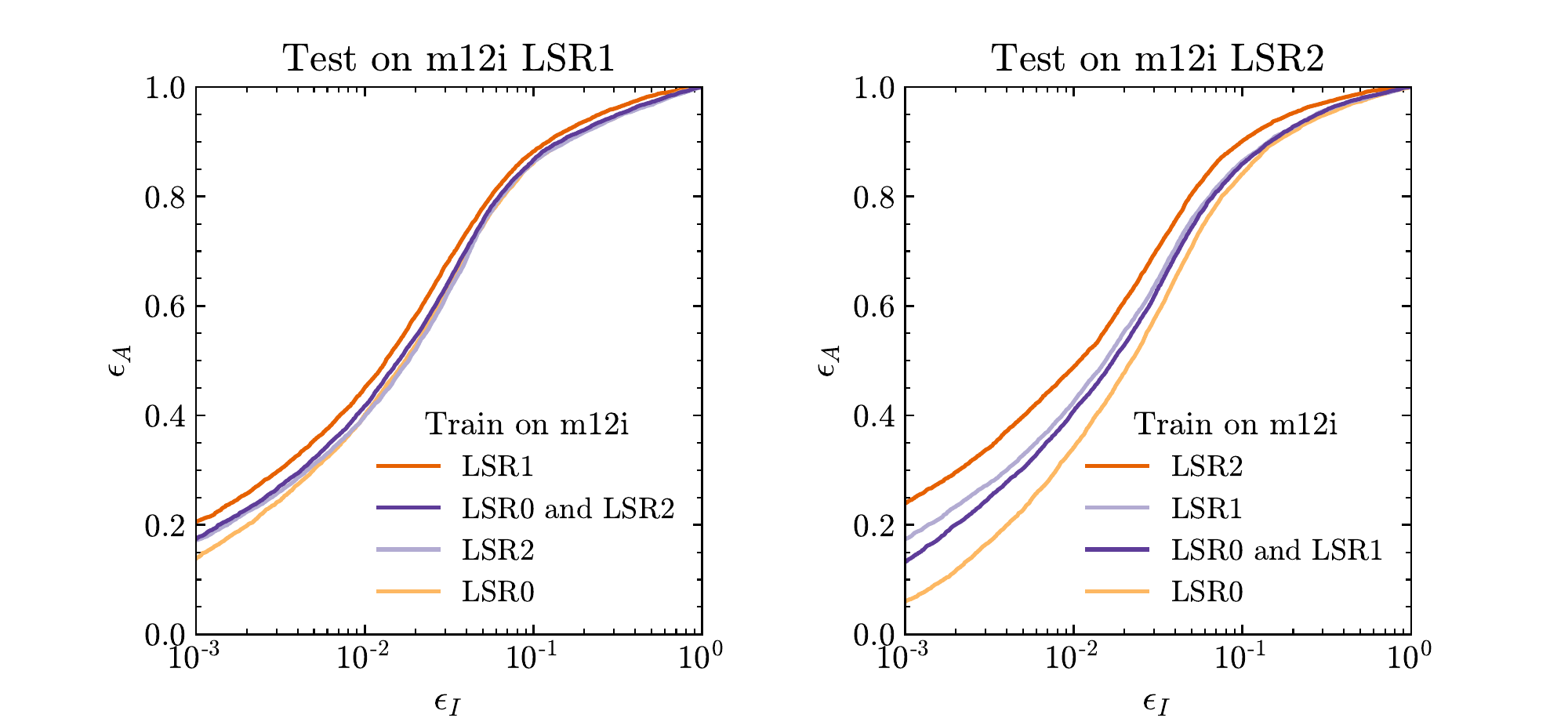}
  \caption{ROC curves showing the performance of networks that are trained on one local standard of rest (LSR) of \mi and applied to a different LSR of \mi.
  The networks are trained and tested on the subset of stars that have small parallax errors, and only use 5D kinematics as inputs.
  The orange line shows the result of training and testing on the same LSR; this provides a benchmark for the best that the network can do.
  In cases where the network is trained on a different LSR than which it is tested, the performance is not dramatically worsened.
  This suggests that the network is learning gross features that distinguish accreted from \insitu stars, rather than detailed features of a specific galaxy.  
  }
  \label{Fig:MultipleLSRKinematics}
\end{figure*}

We train new neural networks on each of the LSRs.
The plots in \cref{Fig:MultipleLSRKinematics} show the resulting ROC curves derived from a number of different networks that are only trained on the 5D kinematic information for each of the three LSRs and a combination of the two not being used for testing.
Each network is then tested on the LSR1 and LSR2 datasets (left and right panels, respectively).
In both panels, the orange line (top in the legend) shows the performance that can be achieved using each local training set and applying it to the same LSR.
This serves to benchmark the performance of the various networks as it allows our networks to see local kinematic information during training.
The yellow lines (bottom in the legend) depict the networks that are trained on LSR0 and applied to either LSR1 (left panel) or LSR2 (right panel); the lavender lines in the left (right) panel depict the networks trained on LSR2 and applied to LSR1 and vice versa, respectively.
In both cases, the lavender ROC curves out-perform the yellow, showing that LSR1 and LSR2 are more similar to each other than to LSR0.

The first lesson we learn from these figures is that the networks are relying on distinguishing features that are universal to all three LSRs, which is why even the worst curve (in yellow) still does a good job rejecting background.
For the LSR0 curve, at a fixed false positive rate $\epsilon_I$ of 0.01, the accreted efficiency $\epsilon_A$ is 0.40 (0.34) for the left (right) plot.
As a comparison, the \textbf{VM selection} had a false positive rate of 0.017 and selected accreted stars at a rate of 0.89 --- but it requires both 3D velocities and metallicity, which are not necessarily present in this data set.
On the other hand, the loss of performance experienced by the networks trained on stars from the same catalog (the orange lines) show that the networks are taking advantage of some additional detailed structures that does not appear in all frames of reference.

Since we want to keep our neural network from learning too many specific details of any particular simulated reference frame, we choose to train it on multiple mock surveys simultaneously.
The purple curves in \cref{Fig:MultipleLSRKinematics} display the result of this test.
In the left panel, we see that the network trained on both LSR0 and LSR2 behaves similarly to that trained only on LSR2, which is better than the one only using LSR0.
In the right panel, we see that the combined network does not perform as well as that using LSR1, but does do better than that using LSR0.
However, in both cases the performance of the combined network is closer to the higher-performing viewpoint extrapolation.
This is encouraging, since a priori we have no way of knowing which choice of training sample will give better performance.
In comparison to training on the worse-performing LSR0 alone, at a false positive rate of 0.01 the combined network gives a gain in the accreted efficiency of 9\% (19\%) for the left (right) plot.

We conclude that training on multiple mock catalogs both raises the baseline performance and helps the networks to focus on general features to distinguish accreted from \insitu stars without overemphasizing particular local substructure.  
Two comments are in order.
First, we remark that all of the examples explored here were taken from the same simulated galaxy.
One might be concerned that the network is still learning specifics of \mi, which are common among the different LSR catalogs (even though the stars themselves only fall into one catalog, since we have restricted the distance from the mock satellite to $\lesssim 4\kpc$).
As part of \Sec{Sec:Transfer}, we address this issue by utilizing another simulated galaxy (\mf) and by applying a method known as transfer learning.
The second comment is that there is a downside to having an overly general network, since the ideal machine would know about detailed properties of the Milky Way.
We at least partially accommodate this desire by bootstrapping our training with actual Milky Way data in order to let the machine learn and subsequently expose the local substructure present in the \Gaia data.

\section{What the machine is using to classify}
\label{sec:WhatIsLearned}

The results presented thus far make it clear that the neural network can distinguish between accreted and \insitu stars. 
This section presents an analysis that utilizes the ``data planing'' method~\citep{Chang:2017kvc} to suggest what information is important for classification.

The power of neural networks is that they can learn complex representations of the data from low-level inputs.
The goal of this technique is to expose the high-level variables being utilized by the machine.
To this end, we construct a ``planed'' dataset by removing high-level information from the original data.
One can then train a new network on the planed data, with the drop in resulting network performance as a quantitative measure of the importance of that variable for the original classifier.
An example of a high-level variable that will prove to carry a lot of discriminating power is the rotational velocity of the stars in the galactic frame.  
We note that the network must learn something analogous to both variable estimation and coordinate transformation on the 5D kinematic inputs to take advantage of this information, since the velocity measurements provided are neither complete nor in cylindrical coordinates.

When planing the data, we suppress information by first generating a histogram in the variable of interest for the accreted and \insitu stars separately.
Each star is then given a weight inversely proportional to the probability that it falls in a particular bin.
These weights are applied to each sample within the loss function during training, which is analogous to uniformly sampling over the given variable.  
Therefore, when we plane in, \eg the rotational velocity, the network should not see the peak associated with the stars rotating with the disk.

\begin{figure}[t]
  \centering
  \includegraphics[width=0.95\linewidth]{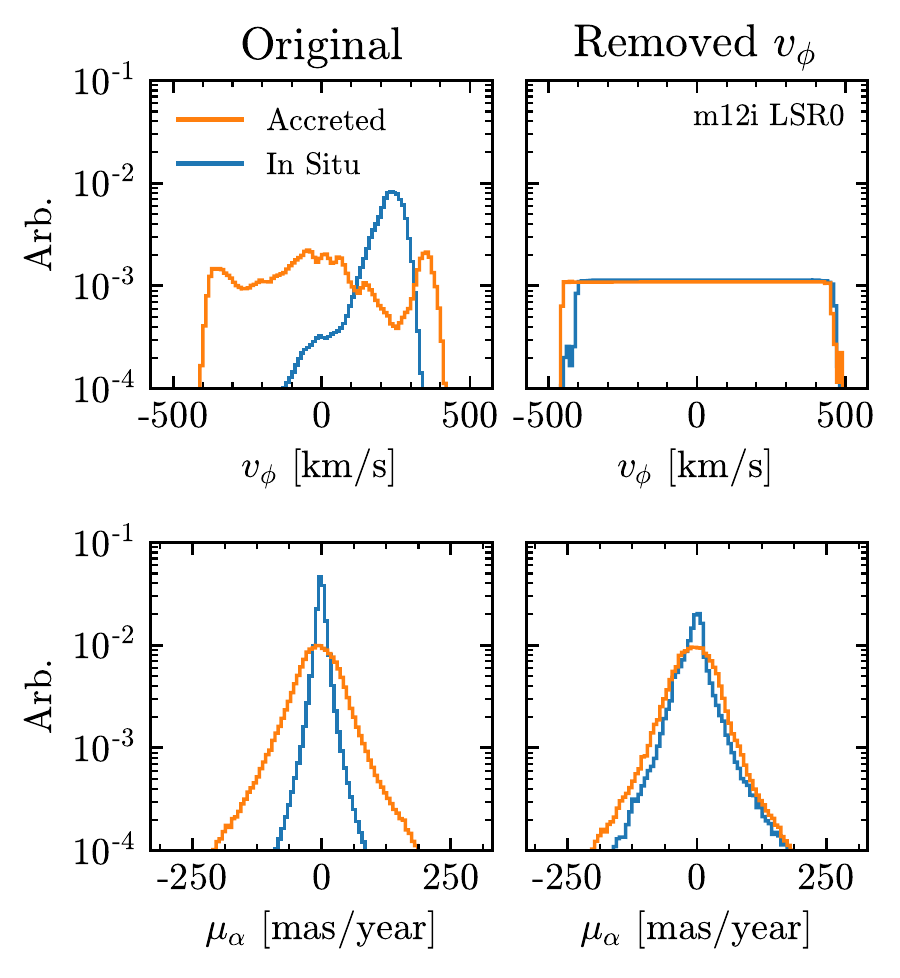}
  \caption{Data planing allows one to determine which high-level variables play an important role in the network classification.
  In this procedure, one starts with the initial distribution (in arbitrary units) of a particular variable, in this example, the rotational velocity $v_\phi$ (top left), and then suppresses the information by weighting each star inversely by the probability that it falls within a particular histogram bin (top right).
  The planing procedure effectively removes the $v_\phi$ information from the dataset, but also affects the distribution of other variables.
  For example, the bottom row shows the proper motion in the right ascension, $\mu_\alpha$, distributions before and after planing in $v_\phi$.
  By running the networks on the planed datasets, one can estimate the importance of the planed variable in driving the classification.
  The distributions in this figure pertain to the subset of \mi LSR0 with small parallax errors and a measurement of $v_{los}$.
  }
  \label{Fig:Planing}
\end{figure}

Any observable that is correlated with the planed variable will have an altered distribution with respect to the original data set.
\Cref{Fig:Planing} shows an example of the planing process on \mi LSR0 for the stars with $\delta\varpi /\varpi < 0.10$ and a $v_{los}$ measurement.
The left column displays the initial distributions of $v_\phi$ and the proper motion in the right ascension direction ($\mu_\alpha$) on the top and bottom panel, respectively.
The distributions after weighting the samples  in inverse proportion to the $v_{\phi}$ probabilities are shown in the right column.
By design, the $v_\phi$ distributions in the planed data set are uniform, up to statistical noise.
The lower right plot for $\mu_\alpha$ shows that uniformly sampling the stars over $v_\phi$ impacts the distributions of the input variables.

Planing the data results in poorer network performance.
To quantify this, we use the area under the ROC curve (AUC) as a simple indicator of a network's ability to separate signal from background.
An AUC of 1 implies that every star is perfectly identified, while an AUC of 0.5 is equivalent to a random guess.
By comparing the results of networks trained on different planed datasets, we can assess how important different observables are for the classification.
The more important the planed variable, the larger the reduction of the AUC.

The AUC of networks trained on different planed datasets is shown in \cref{Tab:Planing}.
We choose to use Galactocentric cylindrical coordinates $(v_R, v_\phi, v_z)$ as the high-level variables, as this reflects the symmetry of the disk.
Using the full dataset (without planing), the base AUC score is 0.96.
Removing the $\phi$ positional information of the stars actually slightly improves the score, presumably because this transformation of the data simply enforces the underlying cylindrical symmetry of the disk.  
Within the solar neighborhood, we also do not expect the galactic distance to play a large role, and we see that removing this only slightly changes the performance.
For contrast, since the \insitu stars are clustered around $z = 0\kpc$ (the disc), planing in $z$ reveals that this variable is somewhat important to the network.
Interestingly, even though the network has access to the 3D position and only 2D velocity, evenly sampling over any of the velocity components reduces the performance more than removing position.
As anticipated, this shows quantitatively that the rotation of the disk is the most significant variable for classifying the accreted stars.

\begin{table}[t]
\begin{center}
    \renewcommand{\arraystretch}{1.5}
	\setlength{\tabcolsep}{30pt}
	\setlength{\arrayrulewidth}{.3mm}
	\begin{tabular}{l c}
    	Data & AUC \\
	\hline\hline
	Using full set & 0.96 \\
	Removed $\phi$ &  0.97\\
	Removed $R$  & 0.94 \\
	Removed $z$ & 0.90 \\
	Removed $v_z$ & 0.84 \\
	Removed $v_R$ & 0.83 \\
	Removed $v_\phi$ & 0.75 \\
	\end{tabular}
\caption{Area under the ROC curve (AUC) for networks trained on different planed datasets of the \mi~LSR0 catalog, and using only 5D kinematics. 
The extent to which the AUC is lowered when removing information represents how important that variable is in distinguishing accreted stars.  For comparison, we also provide the AUC for the original (un-planed) data.  Of the simple kinematic variables considered, the disk rotation is the most crucial, lowering the AUC from 0.96 to 0.75.  This suggests that, even though the machine is not given enough information to fully reconstruct $v_\phi$, it is still inferring a set of correlations of the input variables that correspond to this quantity.}
\label{Tab:Planing}
\end{center}
\end{table}

Even though the networks are not using the information shown in \cref{Tab:Planing} directly, these results hint at what the machine is learning.
Specifically, there is more discriminating power in the kinematic distributions than in the position information.
This additionally explains how the networks are able to generalize so well when testing on different LSRs in a specific galaxy: $v_\phi$ does not vary dramatically among them.
The dominant importance of information about velocities
within the plane of the galactic disk is precisely what one would
expect from our current understanding of galactic dynamics.

We note that while $v_{\phi}$ is the most important variable considered here, it does not reduce the AUC to 0.5, reflecting that the machine is relying on additional information.
Correlations among the variables considered here are likely important.
However, planing is difficult in multiple dimensions, especially when the underlying distributions are not smooth, and is beyond the scope of this work.

\section{Transfer Learning on \FIRE}
\label{Sec:Transfer}

In the last section, we showed that neural networks are able to take advantage of non-trivial kinematic correlations among the 5D inputs to distinguish stars accreted onto a galaxy versus those born within the galaxy.
For instance, even though the algorithm is not given $v_{\phi}$ directly --- there is in fact not even enough information available to fully reconstruct $v_{\phi}$ --- it still infers that the combination of observables roughly corresponding to $v_{\phi}$ is an important quantity.
However, all of these results are based on a study of a single simulated galaxy, and so one may be concerned that the kinematic features being used to discriminate will not be relevant for other galaxies.
In this section, we will demonstrate that this worry is unfounded, in that the network's performance is largely maintained when applying it to a different galaxy with a distinct merger history.
To account for the differences between galaxies, we will use of a scheme known as ``transfer learning" to refine a network's performance on the particular galaxy of interest.
Our implementation of this approach requires initial training on a simulated galaxy, followed by additional restricted training on (simulated or real) data from another galaxy.

\begin{figure*}[t]
  \includegraphics[width=\linewidth]{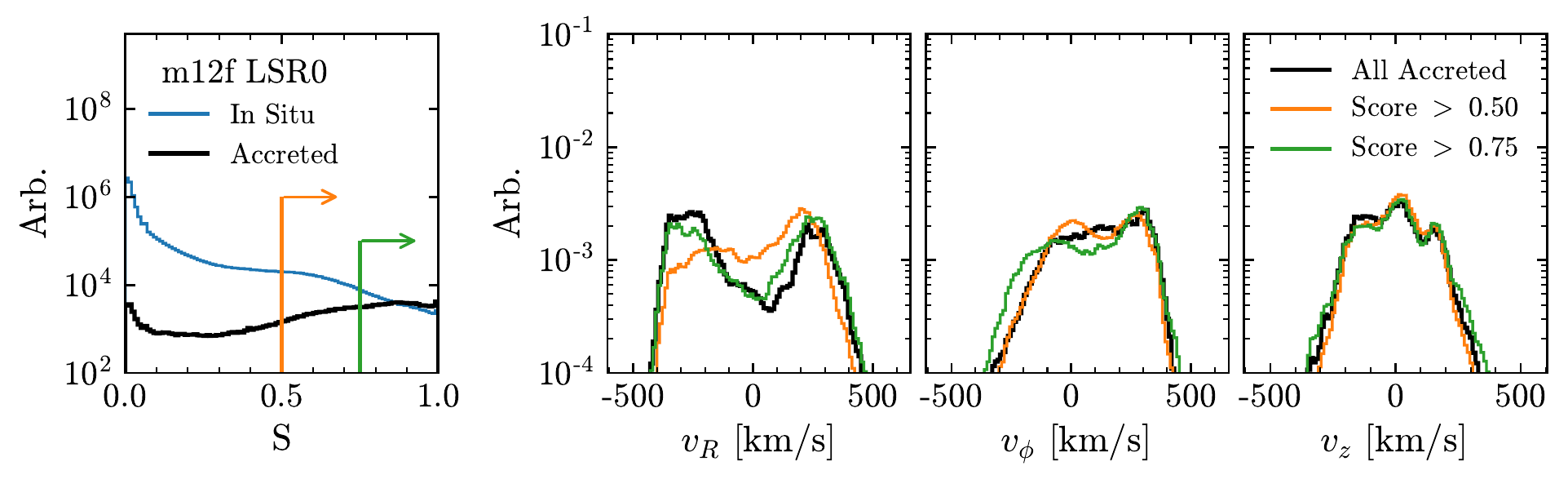}
  \caption{Example metrics for transfer learning.
  The left panel shows the distribution in arbitrary units of the network scores for the true accreted (black) and \insitu (blue) test stars of \mf LSR0.
  We note that the stars that are truly \insitu have network scores peaked towards 0, while those that are truly accreted have scores peaked towards 1.
  The orange and green arrows indicate two cuts on the network scores: $\NN > 0.50$ (orange) and $\NN >0.75$ (green).
  For the  example illustrated here, the scores are specific to a network where transfer learning was performed on the last layer using the \textbf{ZM selection} to derive labels.
  In the remaining panels, the orange and green lines show the normalized $v_R$, $v_\phi$, and $v_z$ distributions of the stars with scores larger than the indicated cut.
  The thick black lines correspond to the distribution of the truth-level accreted stars, not just the ones passing the cuts.
  We see that cutting on a network score of $0.75$ (green lines) better reproduces the truth distributions.
  To quantify the goodness-of-fit, we calculate the $\chi^2$ for the $v_R$, $v_\phi$, and $v_z$ distributions separately, and sum them together to get a total $\sum \chi^2$.
  The lower the value of $\sum \chi^2$, the better the network reproduces the truth distributions.
  For the case illustrated here, the $\sum \chi^2$ is a factor of 3 smaller for the green distributions, compared to the orange ones.}
  \label{fig:Chi2Example}
\end{figure*}

\subsection{Transfer Learning Methodology}
\label{sec:TransferMethodology}
Transfer learning refers to a training strategy whose goal is to utilize a model trained for one task (perhaps where there is a large high quality data set) to perform a different, albeit related, task~\citep{Caruana:1994:LMR:2998687.2998769, Bengio:2011:DLR:3045796.3045800, pmlr-v15-bengio11b, 2013arXiv1310.1531D, NIPS2014_5347}.
The utility of this approach has been demonstrated for image recognition tasks, where the idea is that certain layers of a deep network have learned to recognize specific features in an image, for example, eyes or legs.
However, a network that was trained on many images of high-quality stock photos may not work well for blurry, poorly-lit user images, even though the objects to be identified have many features in common with the stock photos.
Thus, instead of initializing the ``blurry'' network with random weights, the paradigm of transfer learning starts with the weights determined by training on the stock photos.
This procedure is often referred to in the literature as ``pre-training."
Ideally, this implies that the network starts its second round of training near the minimum of the loss function, such that only minor tweaks are needed to optimize for the new data.

Our network will be pre-trained on simulated galaxy catalogs (in analogy with the stock photos), which are similar to the Milky Way.
By training on multiple catalogs, we will obtain a network that is only sensitive to general dynamical features.
Then, transfer learning will be performed on a separate data set from another simulated galaxy.
This will result in a network that can accurately classify accreted and \insitu stars between two galaxies with different merger histories.

\subsection{Transfer Learning Experiments}
\label{sec:TransferExperiments}
To choose the best method of transfer learning for our particular application, we perform the following set of numerical experiments using \mi as the basis for the pre-training, and \mf as the mock ``real'' galaxy.
As we have chosen to use 5-layer networks, there are multiple ways in which the weights of the different layers can be unfrozen to allow the network to specialize.
We consider the following possibilities:
\begin{enumerate}[leftmargin=1.3\parindent]
	\item \textbf{Update only the first layer.} This assumes that the generic network has already learned most of the relevant features, but that the input normalizations vary from catalog to catalog.
	Note that before feeding the data to the networks, the inputs have been normalized by the mean and variance of the \mi LSR0 data.
	By re-training the weights of the first hidden layer, the network can learn the normalizations specific to another galaxy.
	\item  \textbf{Update only the last layer.} This option still assumes that all of the learned features are essentially the same, but that the final combination of these features that ultimately yields the optimal classification is different for the new galaxy.
	Re-training the network fixes the combination of high-level features learned in the pre-trained network.
	\item  \textbf{Update the first and last layers.} We also consider a combination of options two and three.
	\item  \textbf{No transfer learning.}  As a means of comparison, we provide results where the pre-trained network is used directly without updating any of the weights.
\end{enumerate}

There is an additional aspect of the method that requires testing.
The stars in the new galaxy that are used for re-training must be given a label.
While we have access to truth-level information in the simulations, we obviously do not for the Milky Way.
Therefore, we also test which of the traditional cut-based methods works best as a way to derive such labels.
This allows us to evaluate whether it is better for the network to see more examples of truly accreted stars (while inherently mislabelling more \insitu stars as accreted), or to use stars that have a much higher probability of being accreted (while incorrectly labeling many accreted stars as \insitu).

We take the \mf LSR0 catalog to be our ``real'' data set.
The initial pre-training uses a combined dataset of the LSR0, LSR1, and LSR2 catalogs of \mi for stars with $\delta\varpi/\varpi < 0.10$, providing us with an initial training set containing 176,842,422 \insitu stars and 1,427,323 accreted stars.
In the transfer learning step, we use 200,000 randomly drawn stars from the \mf LSR0 catalog with $\delta\varpi/\varpi < 0.10$ and measured radial velocities.\footnote{This sample size was chosen to be comparable to the RAVE~DR5-\Gaia~DR2 catalog (see \cref{tab:StarNumbers}), which is the data set that will be used when ultimately performing transfer learning for the Milky Way.}
The initial learning rate for the transfer learning training step is set to $10^{-4}$.
Although the results are similar when using a larger initial learning rate, the overall performance suffers.  The resulting networks derived for all the different transfer learning options are compared against a test set of 10,000,000 stars pulled from the larger \mf~LSR0 dataset, \ie, these stars are not required to have a line-of-sight velocity measurement.

\begin{figure}[t]
    \centering
    \includegraphics[scale=0.9]{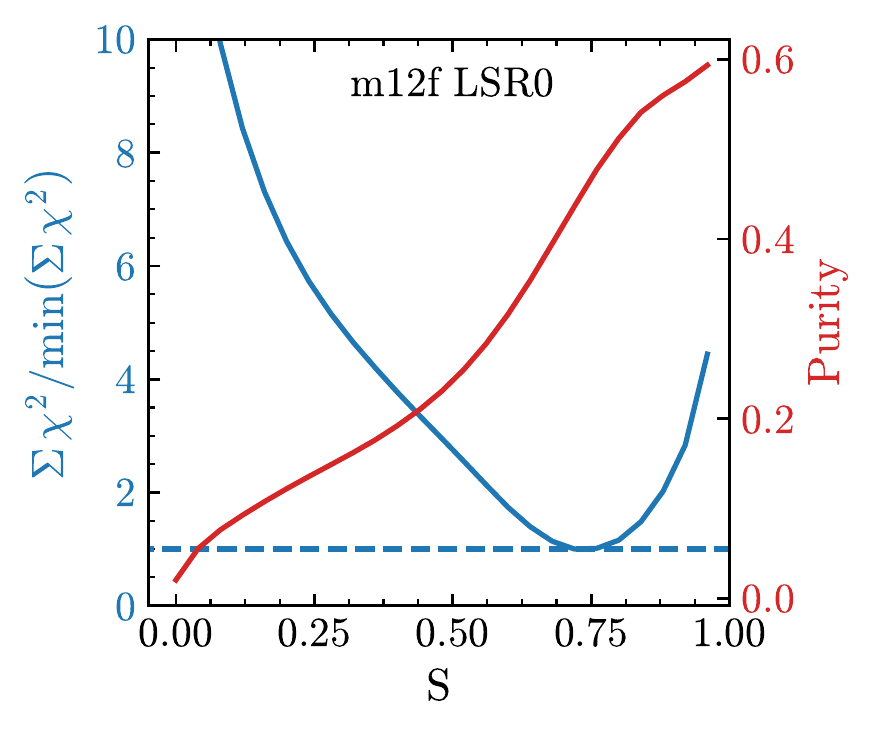}
    \caption{Determination of the transfer learning method. The blue curve shows the sum over the $\chi^2$ of the normalized $v_R$, $v_{\phi}$, and $v_z$ distributions compared to all of the accreted stars in the test sample, normalized to the minimum value obtained. The $x$ axis shows the cut value, where all stars with a network output larger than this are selected. The network was trained with transfer learning updating only the last layer and used the \textbf{ZM} labels. A cut of 0.75 gives the best overall fit. The red curve shows the purity of the resulting sample (and corresponds to the labels on the right axis). Tighter cuts on $\NN$ result in a more pure sample, but may bias the resulting kinematic distributions when stronger than $S\sim 0.75$.
    }
    \label{fig:sumchi2}
\end{figure}

We need a metric to evaluate which transfer learning approach is optimal.
To this end, we compare the recovered distributions for the 3D velocities ($v_R, v_\phi, v_z$) to the truth-level distributions.
The left-most panel of \cref{fig:Chi2Example} shows the network score for the truth-level accreted (black) and \insitu stars (blue) for one example.
In this case, the training labels for the \mf LSR0 set used for transfer learning were determined using the \textbf{ZM selection} and transfer learning was only performed on the last layer of the network.
As desired, the network scores are peaked towards 1 for the truly accreted stars, whereas the \insitu stars have scores peaked towards 0.
The remaining panels in \cref{fig:Chi2Example} show the normalized velocity distributions, where the thick-black line is the truth-level distribution for the accreted stars.
The orange and green lines show all of the stars from the test set (not just the accreted stars) that have network scores greater than 0.5 and 0.75, respectively.
The orange distributions do not match those of the accreted stars, yielding $\chi^2$ values of $2.46\times10^{-2}$, $2.12\times10^{-3}$ and $1.54\times10^{-3}$ for $v_R$, $v_{\phi}$, and $v_z$, respectively, where $\chi^2$ is defined as
\begin{equation}
\chi^2 = \sum_{i\in \rm{bins}} \big(f(v)_i - h(v)_i \big)^2,
\end{equation}
and $f(v)_i$ is the value of the distribution of the truly accreted stars in velocity $v$ bin $i$ and $h(v)_i$ is the same except for the stars passing the neural network cut.
In comparison, the $\chi^2$ values for the green distributions are $4.38\times10^{-3}$, $4.77\times10^{-3}$, and $1.07\times10^{-3}$.
A cut on the network score of 0.75 yields a lower sum of these $\chi^2$ values, $\sum \chi^2$, by about a factor of 3.
This suggests that cutting on a network score of 0.75 yields distributions that are closer to truth; this is clear from \cref{fig:Chi2Example} where the green lines trace the black better than the orange lines do.

This procedure suggests an operative definition of an optimal value for the cut on $\NN$, as shown in \cref{fig:sumchi2}.
The blue line corresponds to the left axis, and shows $\sum \chi^2$ relative to the minimum value. 
The red curve (correspond to the right axis) shows the purity of the sample after making the cut.
A cut of 0.75 minimizes $\sum \chi^2$, while a tighter cut results in a more pure sample.

\begin{table}[t]
\begin{center}
    \renewcommand{\arraystretch}{2}
	\setlength{\tabcolsep}{10pt}
	\setlength{\arrayrulewidth}{.3mm}
	\small
    \begin{tabular}{c | l  c}
         & Trainable layers & $\sum \chi^2~~[\times10^{-2}]$\\ 
        \hline\hline
        & No transfer learning & $1.07$ \\  
        \hline
        \multirow{3}{*}{\rotatebox[origin=c]{90}{\parbox[c]{1.0cm}{\textbf{VM}}}}
        & 1st &  $2.13$ \\ 
        & Last & $1.20$ \\  
        & 1st and last & $2.28$ \\  
        \hline
	\multirow{3}{*}{\rotatebox[origin=c]{90}{\parbox[c]{1.0cm}{\textbf{ZM}}}}
        & 1st & $1.04$ \\  
        & Last & $1.02$ \\ 
        & 1st and last & $1.03$ \\  
    \end{tabular}
    \caption{Results for a neural network trained on LSR0, LSR1, and LSR2 of \mi using only 5D kinematics as inputs.  We compare the results with and without transfer learning on \mf LSR0.  When the transfer learning is performed, either the first layer is updated, or the last layer is updated, or both.  In the transfer learning step, all truth labels are derived using either the \textbf{VM} or \textbf{ZM selections}, as defined in \Sec{Sec:CutAndCount}.  $\sum \chi^2$ quantifies how similar the $v_R$, $v_\phi$, and $v_z$ are to the truth-level distributions, as illustrated in \cref{fig:Chi2Example}.  For each transfer learning method, we determine the optimal cut on the network score, as described in the text.  In general, the \textbf{ZM selection} with transfer learning performed by only varying the last layer does the best job at reproducing the kinematic distributions of the true accreted stars, \ie, it achieves the lowest $\sum \chi^2$.}
    \label{Tab:TransferKinematicResults}
    \end{center}
\end{table}

We repeat this procedure for each transfer learning and empirical labeling method, finding the optimal cut value on the network score using $\sum \chi^2$ in the same way for each case.
The optimal cut value typically falls between 0.5 and 0.9 for each method.
This optimal cut is not always close to unity, since although increasing the value of the cut increases the fraction of stars that are accreted in the final selection, doing so also biases the distributions.
In other words, the selection of stars that are most easily identified as accreted do not reproduce the underlying full distribution of accreted stars.

\begin{figure*}[t]
    \centering
    \includegraphics[width=0.75\textwidth]{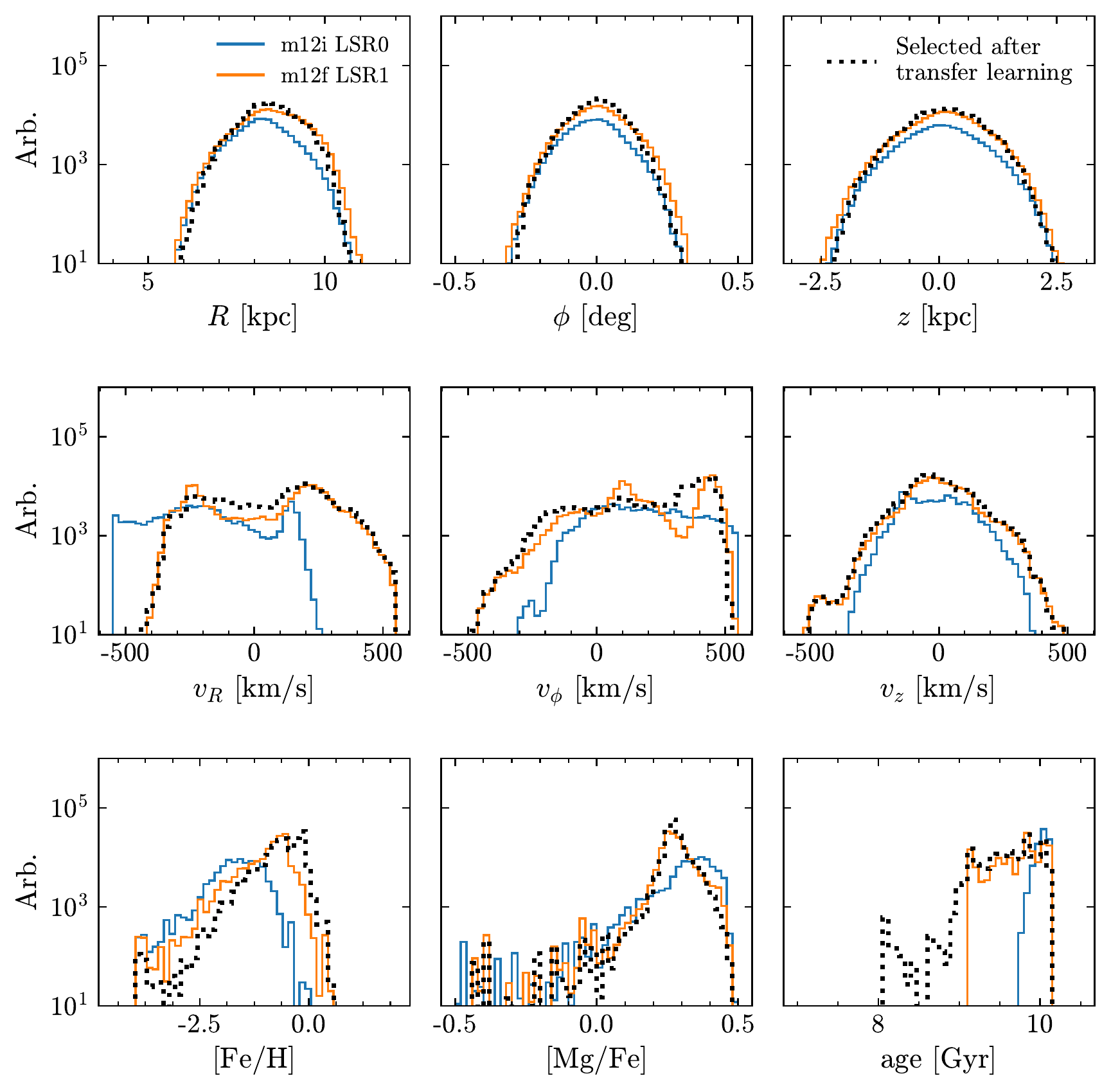}
    \caption{Validation of the transfer learning procedure. All distributions are shown in arbitrary units.
    The network is pre-trained on \mi LSR0 using 5D kinematics as inputs.  The true 6D distributions of accreted stars that the network sees are depicted by the blue lines in each panel.  Then, transfer learning is performed by updating the last layer of the network on \mf LSR1 using 200,000 stars that are labeled using the \textbf{ZM selection}; that is, they are labeled as accreted if $|z| > 1.5\kpc$ and $\FeH <-1.5$.  Stars from \mf LSR1 that have a network score greater than 0.75 are then marked as accreted.  Their distributions are indicated by the black dotted lines in the panels.  For comparison, we also show the truth-level distributions for accreted stars in \mf LSR1 as the orange lines.  The distributions of the stars selected by the network do an excellent job at reproducing the truth-level distributions for the accreted stars in \mf, even though the network was pre-trained on an entirely different galaxy with a different merger history.  This result justifies our confidence that a network trained on simulation can be successfully applied to the Milky Way.
    }
    \label{Fig:MockAnalysis}
\end{figure*}

\Cref{Tab:TransferKinematicResults} shows the benchmark performance for the different networks tested here through their minimum resulting $\sum\chi^2$ values.
Interestingly, using the \textbf{VM selection} as the transfer learning label always results in worse fits.
This can be traced to the fact that the \textbf{VM selection} has large contamination from \insitu stars, thereby negatively affecting the results.
In addition, as this selection uses velocities (which as shown above are more important than spatial information), it has more potential to create a biased set. 
In contrast, the \textbf{ZM selection} label yields lower $\sum \chi^2$ values, likely because of the low false-positive rate of this method.
As a result, we choose to use the \textbf{ZM selection} and only update the last layer the network.\footnote{It could be argued that performing transfer learning to get a 5\% better fit is unnecessary, especially when it relies on using empirical labels to define ``truth.'' However, we will find that the use of transfer learning is a critical step towards obtaining a reliable catalog of real Milky Way stars. This will be discussed more in \Sec{sec:Catalog} where the actual catalog is presented.}
Furthermore, we take the optimal cut on the neural network output to be 0.75 so as not to bias the velocity distributions --- any star with a network output greater than this will be labeled as accreted.

\subsection{Transfer Learning Validation}
\label{sec:mock}

The methodological decisions in our transfer learning regimen and the determination of the optimal cut value on the neural network output were justified by pre-training networks on \mi and then transferring and testing on \mf LSR0.
The goal of this section is to validate these choices.
In particular, we will apply the machinery of the previous section to a different viewpoint within \mf, specifically LSR1.  
Truth-level information will only be utilized for computing performance metrics.

As before, we begin with the networks pre-trained on \mi.
Next, we mock the dataset used for the transfer learning training step by randomly selecting 200,000 stars from the \mf LSR1 catalog with the requirement that they have $\delta \varpi/\varpi < 0.10$ and a radial velocity measurement.
Stars within this subset that have $\FeH <-1.5$ \& $|z| > 1.5\kpc$ are assigned a label of 1, and the stars that do not pass this cut are labeled with 0.
The inputs are rescaled according to \cref{Tab:Normalizations}.
The last network layer is then unfrozen and transfer training is performed.

The results are presented in \cref{Fig:MockAnalysis} for all test stars in \mf LSR1 with $\delta\varpi/\varpi < 0.10$. Note that we are able to plot the 6D information since we have access to the truth-level phase space.
The blue lines are the distributions of the accreted stars in the \mi LSR0 catalog, depicting the shapes on which the network was pre-trained.
The orange lines show the truth-level distributions for the accreted stars in \mf LSR1; these clearly differ in important ways from the \mi LSR0 distributions.
The black, dotted lines display the stars from the \mf LSR1 catalog selected after the transfer learning.
Importantly, the distributions of the stars that are identified as accreted by the network closely trace the truth-level distributions of \mf LSR1.
This demonstrates that the network after transfer learning can successfully select accreted stars from a galaxy that is not the same as the one used in pre-training, even though the distributions of the two galaxies have distinct features.
It is a little troubling that the network does classify some younger stars as accreted; however, this is a small fraction of the selected stars.
As the network is only using kinematic information, it is hard to distinguish what is causing this selection.

Of the 10 million stars in the test set, only 163,727 are accreted.
When using the cut of 0.75, there are 187,721 stars selected, of which 77,614 are accreted.
The purity of the selection is 41\%, comparable to the standard methods in \Sec{Sec:CutAndCount}, while the accreted selection efficiency is 47\%.
The cut value can be increased to get a more pure sample at the risk of biasing the distributions. If the cut is placed at 0.95 then there are only 35,495 stars selected, with 20,897 being accreted.
This brings the purity to 59\%.
While this results in a selection efficiency of only 13\%, this is better than any of the methods in \cref{Tab:TradMethods}.
These results are summarized in \Cref{Tab:MockAnalysis}.

\begin{table}[t]
\begin{center}
    \renewcommand{\arraystretch}{1.7}
	\setlength{\tabcolsep}{8pt}
	\setlength{\arrayrulewidth}{.3mm}
    \begin{tabular}{l c c}
        Data set & Purity & $\epsilon_A$\\
        \hline\hline
        $\NN(\text{star}) > 0.75$ & 41\% & 47\%\\
        $\NN(\text{star}) > 0.95$ & 59\% & 13\% \\
    \end{tabular}
    \caption{Test set of the \mf LSR1 catalog using stars with $\delta\varpi/\varpi < 0.10$. 
    The neural networks are trained on separate \mi galaxy simulations with truth-level information and then re-trained using only observable information from \mf LSR1.
    The optimal cut of 0.75 yields velocity distributions most similar to the truth-level accreted distributions, but only has a purity of 41\%.
    This is comparable to the traditional methods listed in \cref{Tab:TradMethods}.
    A stronger cut of 0.95 increases the purity, but decreases the fraction of accreted stars that are selected, and can bias the velocity distributions.
    This provides an estimate for the purity and efficiency expected when applied to \Gaia.
    }
    \label{Tab:MockAnalysis}
    \end{center}
\end{table}

This analysis has demonstrated that it is possible to classify accreted versus \insitu stars using simulated data that has been processed to very closely resemble \Gaia DR2.
The optimized strategy relies on beginning with a network originally trained on simulated galaxies followed by the use of transfer learning to adjust the network by augmented training on real data.
We conclude that it is reasonable to apply this method to the \Gaia data, where the transfer training data set is taken from stars in \Gaia DR2 that have been cross matched with spectroscopic surveys.
Cutting on the neural network output (at the optimized value) yields a \Gaia DR2 catalog of accreted stars with reasonable purity and unbiased kinematic distributions.
\Cref{Tab:MockAnalysis} acts as an estimate of the performance we expect in the \Gaia catalog.

Finally, we note that with the large sample sizes considered here, there are examples of accreted stars that the network scores very low and also \insitu stars that the network gives high scores.
Unsurprisingly, we find that the accreted stars that the network thinks are \insitu tend to have kinematics very similar to that of the disk.
The \insitu stars that get high scores tend to be either at larger distances from the galactic plane, or have large retrograde velocities.

\section{Deriving the Catalog}
\label{sec:Catalog}

\begin{figure*}[t]
\begin{center}
\includegraphics[width=0.8 \linewidth]{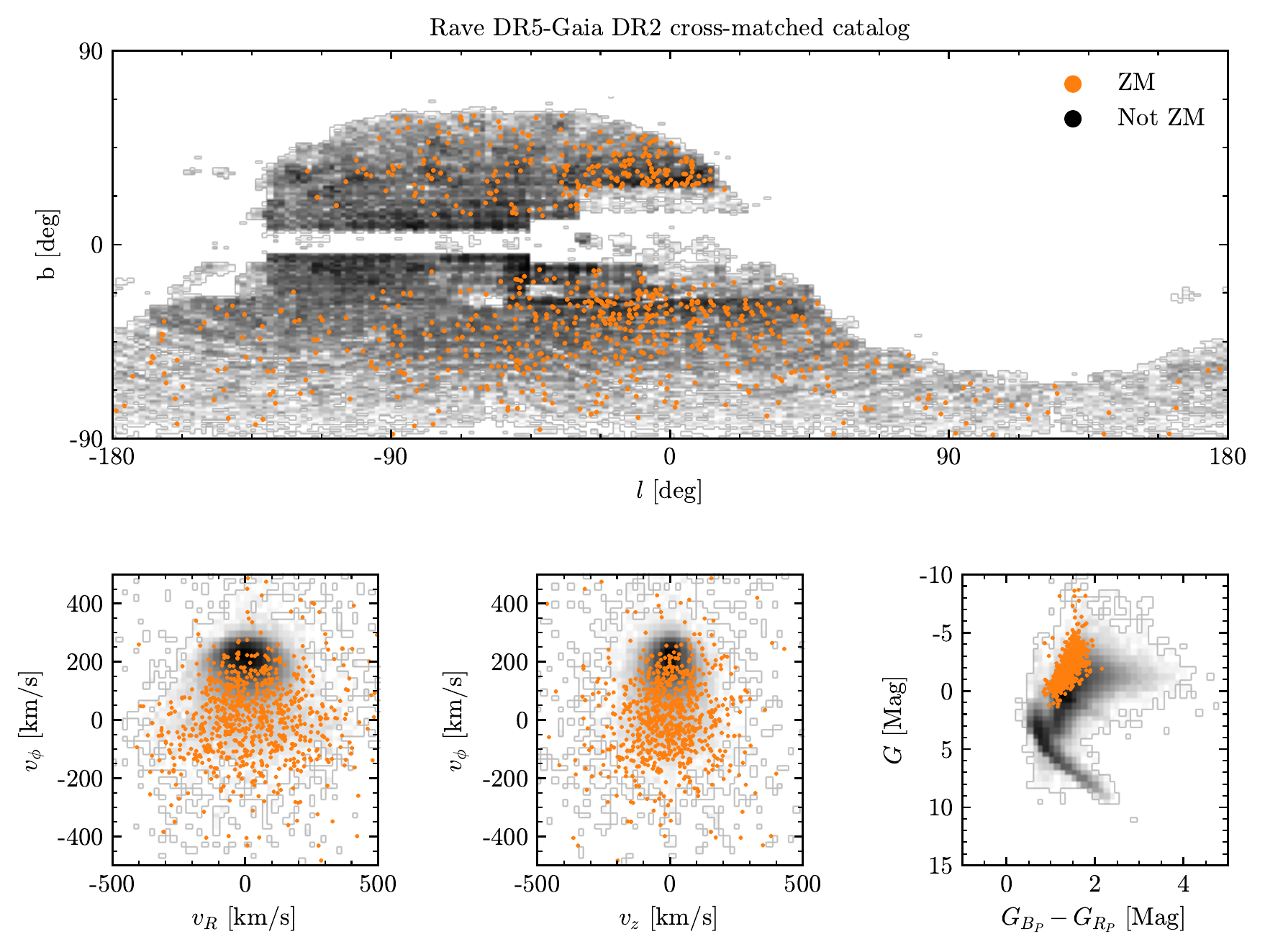}
\caption{Stars used for transfer learning to re-train the networks that were pre-trained on the \mi catalogs from \FIRE. The points marked in orange have $|z| > 1.5\kpc$ and $\FeH < -1.5$ and are labeled as accreted for the training.
The remaining stars are marked as \insitu during the training step.
Due to the large sample size, the stars labeled as \insitu are displayed as a density plot, where the darker bins have a larger number of stars. The HR diagram in the lower right panel does not take into account the extinction and reddening~\citep{2018A&A...616A..10G}.
}
\label{Fig:RaveTraining}
\end{center}
\end{figure*}

Having optimized our approach using mock catalogs, we turn to constructing  neural networks that will be used to distinguish accreted from \insitu stars within the real \Gaia DR2 data.
The number of stars in the \Gaia DR2 release passing the various cuts are shown in \cref{tab:StarNumbers}.
The first stage of training utilizes the combined datasets for the three LSR catalogs of \mi, and takes around 26 hours on dual E5-2690v4 (28 cores) nodes.
Then transfer learning is implemented by updating the weights of the last layer of the network by training on the cross-matched \RAVE~DR5-\Gaia~DR2 catalog \citep{2017AJ....153...75K} and takes around 1 hour.
There are 347,979 stars in the cross-matched catalog that have $\delta\varpi/\varpi < 0.1$, a measurement of the line-of-sight velocity, and have iron abundance.
We label any star that further passes a $|z| > 1.5\kpc$ and $\FeH < - 1.5$ cut as a 1 for training; this cut yields a total of 849 stars.
Stars that do not pass this cut are labeled as 0.
The distribution in the $b$--$l$ plane for both sets of stars are shown in the top panel of \cref{Fig:RaveTraining}.\footnote{We verified that restricting the training stars to follow a similar distribution on the sky does not yield any important effects on the transfer learning. We repeated the analysis of \Sec{sec:mock} using a selection function similar to \RAVE, while still only selecting 200,000 stars. The final results only changed minimally compared to those of \cref{Fig:MockAnalysis}.}

\begin{table}[t]
    \centering
    \renewcommand{\arraystretch}{1.7}
    \setlength{\tabcolsep}{6 pt}
    \setlength{\arrayrulewidth}{.3mm}
    \begin{tabular}{l r}
        \Gaia~DR2                                 & Number of stars \\
        \hline\hline
        Full set                                  & 1,692,919,135  \\
        $\delta\varpi / \varpi < 0.10$            & 72,488,013 \\
        $\delta\varpi / \varpi < 0.10$; $v_{los}$ & 5,393,495 \\
        $\delta\varpi / \varpi < 0.10$; RAVE~DR5  & 347,979 \\
    \end{tabular}
    \caption{Number of stars in the \Gaia DR2 dataset. Requiring a line-of-sight velocity measurement or metallicity (for example as a cross match with \RAVE~DR5) greatly reduces the number of available stars.}
    \label{tab:StarNumbers}
\end{table}

Some resulting distributions for this training data are shown in the bottom panel of \cref{Fig:RaveTraining}.
Note that while the stars given a training label of 1 are relatively evenly distributed in velocity space (the left and middle panels), they are tightly clustered in the HR diagram (the lower right panel).
We do not take extinction or reddening into account in the HR diagram~\citep{2018A&A...616A..10G}.
Additionally, the $v_{R}$--$v_\phi$ plot makes it clear that some of the training stars that are labeled as accreted have rotations that are consistent with the disk, even though they lie outside of the thin disk (by construction). This is important because it shows the network that accreted stars in the Milky Way can have disk-like rotations.

It is worth noting that the testing on mock data presented in \Sec{Sec:Transfer} did not make an overwhelmingly compelling case that transfer learning is necessary.
However, transfer learning has a much bigger impact when applied to \Gaia DR2.
As a cross-check, we performed the same analysis only using the pre-trained network, without transfer learning on the \Gaia DR2 stars.
Only 0.1\% of the stars were selected as accreted, while the simulations suggest between 0.89--1.6\% of the stars with $\delta\varpi / \varpi < 0.10$ should be accreted. With the transfer learning, we select 1.06\% of the stars.
We interpret this result as implying that transfer learning allows the neural network to ``unlearn'' some hidden correlations among the star kinematics within the simulated datasets.
One such potential effect is that each star particle in the simulation is used as a seed for many stars in the catalog, which could yield some unphysical correlations.

\subsection{Validating on Milky Way Data}
\label{sec:CatalogValidation}

\begin{figure*}[ht]
  \centering
  \includegraphics[width=\textwidth]{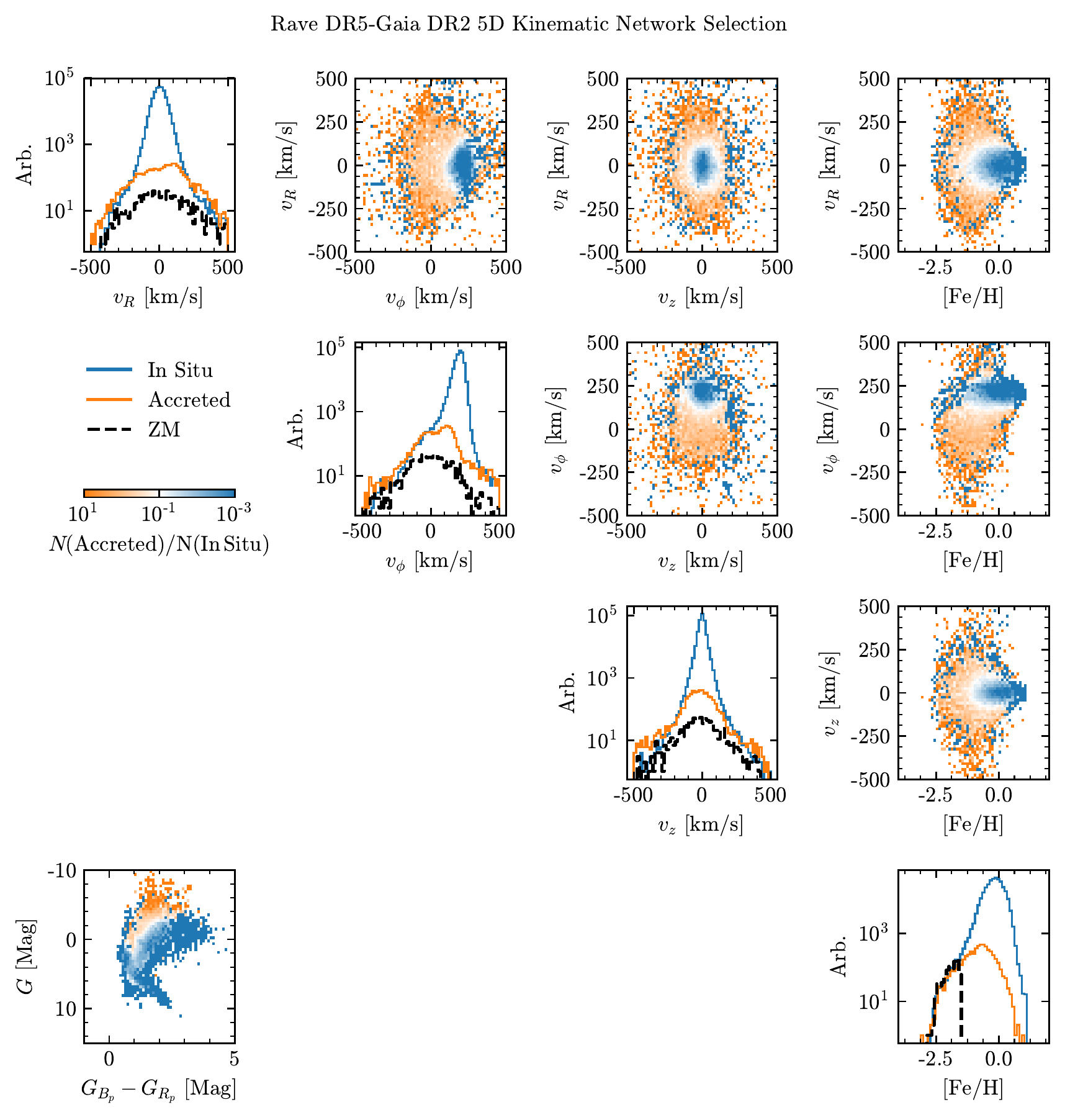}
  \caption{Distributions of the stars in the \RAVE DR5--\Gaia DR2 catalog.
  The blue and orange lines denote stars labeled as \insitu or accreted by the network using only 5D kinematics.
  The black-dashed lines show the distributions of the stars selected with the \textbf{ZM selection} method.
  The stars selected by the network have similar kinematic profiles, but extend to larger metallicities.
  The fact that these distributions agree with the expected halo distributions gives us confidence in our ability to derive a catalog of accreted stars for which we do not have access to a metallicity measurement.}
\label{Fig:RaveKinematics}
\end{figure*}

We perform one final validation of our approach, the results of which are presented in \cref{Fig:RaveKinematics}.
First, we split the cross-matched \RAVE-\Gaia dataset into thirds.
Then we treat one of these thirds of the data as a test set, and we perform the transfer learning by training on the other two thirds.
The resulting network is applied to the test set.
We repeat this procedure twice more by taking the pre-trained network and treating each of the other thirds as the test set.
This allows us to maximize the statistics for validation while ensuring that the network has not seen any of the test stars when training.

As justified above, we classify any star with a neural network output score above 0.75 as accreted, while any star with a lower score is \insitu. 
Some properties of this final validation are provided in \cref{Fig:RaveKinematics}, which shows the resulting 1D distributions along the diagonal, with the stars labeled as \insitu or accreted in blue or orange, respectively.
Additionally, we show the distributions of the stars labeled by applying the \textbf{ZM selection} cuts.
The \insitu stars exhibit a distinct peak in the $v_\phi$ distribution corresponding to the rotation of the disk.
The stars labeled as accreted are roughly symmetric about $0\kms$ in all three velocity components, as expected.
Finally, the \FeH distribution shows that the \insitu stars tend to have larger metallicities than the accreted stars.
Recall that the network is not provided with any metallicity information as an input.
The metallicity distribution of the accreted stars identified by the network extends across the full observed range.

The remaining panels in \cref{Fig:RaveKinematics} show 2D histograms of the ratio of the number of accreted stars to the number of \insitu stars per bin.
The bluest bins have a factor of $10^3$ more \insitu stars than accreted (or they contain no accreted stars).
The orange bins have more accreted stars than \insitu stars, where the largest ratio observed in any given bin is 10.
These distributions reveal that the \insitu stars dominate specific regions of phase space, while the accreted stars span a much larger range.

We interpret \cref{Fig:RaveKinematics} as providing further validation that the kinematic network is working as expected.
This provides confidence that our \Gaia DR2 catalog of accreted stars will be physically meaningful.

\subsection{A Catalog of Accreted Stars}
\label{sec:FinalCatalog}

Now we are finally ready to obtain the neural network used to generate the catalog.
We again start by pre-training on simulation (the mix of all three LSRs within \mi with $\delta{\varpi}/\varpi < 0.10$ and no $v_{los}$ requirements), and then perform transfer learning using the full \RAVE~DR5--\Gaia~DR2 cross-matched catalog as the training data, with labels determined by the \textbf{ZM selection}.
Since we do not separate off a test set, we are not able to perform any validation tests of the final network.
However, this choice has the benefit that it allows us to maximize our statistics during transfer learning to obtain the best possible performance for our ultimate \Gaia DR2 catalog.

\begin{figure*}[t]
\includegraphics[width=\linewidth]{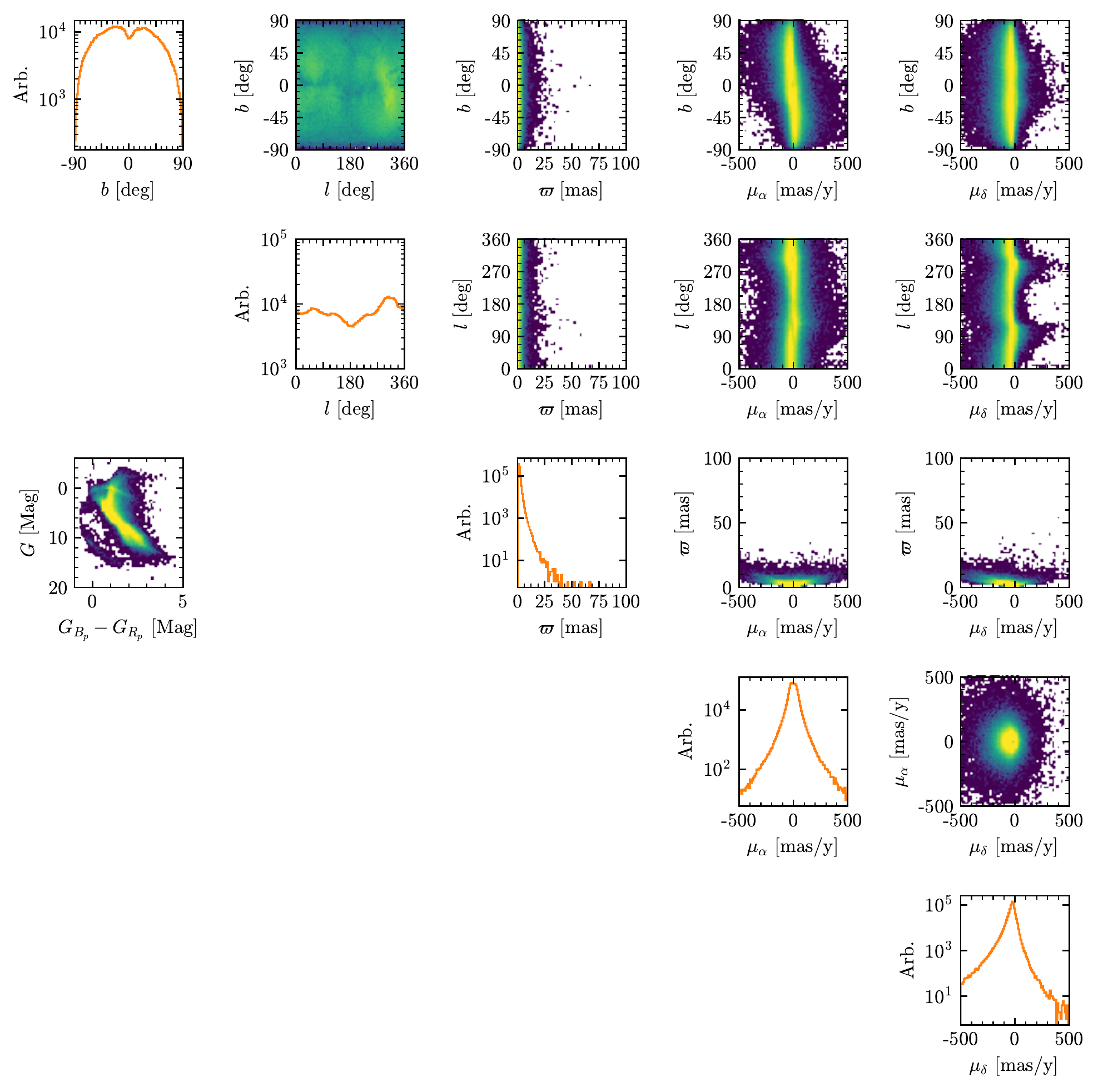}
\caption{Stars within \Gaia DR2 with $\delta\varpi/\varpi < 0.10$.
These stars have been identified as accreted by our neural network using 5D kinematics alone.
While the HR diagram is shown, photometric information is not used in these selections.
As expected for an accreted population, these stars are nearly uniformly distributed in the $l, b$ plane.
See \cref{Fig:InSituKinematicAll} for the analogous figure for the stars classified as \insitu.
}
\label{Fig:AccretedKinematicAll}
\end{figure*}

After this last stage of training, the network is applied to every star within the \Gaia DR2 catalog with $\delta\varpi/ \varpi < 0.10$, which yields $\sim 72$ million total stars, as shown in \cref{tab:StarNumbers}.
Stars with a network score greater than 0.75 are called accreted and added to our catalog.
This yields a total of 766,993 stars that are identified by the neural network as being accreted.
The distributions are shown in \cref{Fig:AccretedKinematicAll}, and the resulting catalog is available on \href{https://doi.org/10.5281/zenodo.3354470}{\textcolor{blue}{Zenodo}}~\citep{zenodo}.
\cref{Fig:InSituKinematicAll} in \App{app:InSitu} provides the distributions of the stars marked as \insitu for comparison.

\section{Conclusions}
\label{sec:Conc}

In this work, we generate a catalog of stars identified by a neural network as having been accreted onto the Milky Way, as opposed to forming \insitu.
Extensive testing was provided to justify a methodology based on transfer learning methods.
One possible limitation of this study is the necessity of training on computationally expensive simulations.
The pre-training was performed on a mixture of simulated mock \Gaia data derived for three different viewpoints within the \mi galaxy from the \FIRE simulations.
The last layer of this network was then unfrozen and allowed to re-train on \RAVE--\Gaia cross-matched data, where the training labels were derived with a traditional cut-based selection (specifically the \textbf{ZM selection}).
The resulting network was applied to the subset of \Gaia DR2 stars with small parallax errors, $\delta\varpi/\varpi < 0.10$.
Stars with a neural network score greater than 0.75 are tagged as accreted.
The resultant catalog is available at \href{https://doi.org/10.5281/zenodo.3354470}{https://doi.org/10.5281/zenodo.3354470}.

The testing performed above used mock \Gaia catalogues generated from cosmological zoom-in hydrodynamic simulations.
Within the simulations, we are able to derive ``truth-level'' labels for the stars using the merger tree history.
We first explored how much information as measured by \Gaia was necessary to accurately classify stars.
This was a critical test since not every star in \Gaia DR2 has been observed accurately enough to provide a complete measurement of its 6D phase space with small errors.
From these explorations, it was found that using 5D inputs --- \texttt{l}, \texttt{b}, \parallax, \pmra, and \pmdec{} --- both provides good classification accuracy and the ability to generalize from nearby stars with full 3D velocity measurements to those that are further away such that $v_{los}$ is unknown.
In addition, we showed how transfer learning can be used to allow a network trained on one simulation to generalize to an independent simulation, using a small amount of local data for re-training.
This local data does not need to have truth-level information, which would be unavailable for a real galaxy, but could be labeled using traditional cut-based techniques to identify accreted stars. 
We note that these traditional techniques can only be used for a small subset of the stars, since they often rely on either knowledge of the full 6D phase space, or spectroscopic measurements that require cross matching between \Gaia DR2 and another dataset.
We showed that using a more restrictive definition of accreted stars for labeling the stars during the transfer training worked better than a more inclusive definition, which justified our choice of the \textbf{ZM selection} (the requirement that $|z| > 1.5$~kpc and $\text{[Fe/H]} < -1.5$) for this purpose.

After identifying the specifics of our transfer learning methodology, we took the network pre-trained on the simulated galaxy viewpoints, and re-trained the network on the \RAVE~DR5--\Gaia~DR2 cross-matched catalog.
The resulting network was then applied to the 72 million stars in the \Gaia DR2 catalog that have parallax measurement errors of less than 10 percent.
Of these, our network identifies 766,993 stars as having accreted onto the Milky Way.
Because the network does not require full 3D velocity information, the number of identified stars is a factor of 21 larger than what would be possible using traditional (3D velocity-dependent) methods.
This large increase in the number of stars allows for a much more detailed study of the halo structure and the history of the Milky Way.
In a companion paper, \citep{catalog_paper}, the catalog is used to reproduce known structures, such as \Gaia Enceladus and the Helmi stream, and identify a new structure.
The new structure is studied in more detail in another companion paper, \citep{nyx_paper}, where we argue the new structure is the remnant of a merged dwarf galaxy. 
Our catalog is a new tool for performing Galactic archaeology that can be leveraged to unlock novel discovery potential thus far hidden within \Gaia DR2.

\section*{Acknowledgments}
We are grateful to Ben Farr and Graham Kribs for useful discussions.  This work utilized the University of Oregon Talapas high performance computing cluster.
BO and TC are supported by U.S. Department of Energy (DOE), under grant number DE-SC0011640.
LN is supported by the DOE under Award Number DE-SC0011632, and the Sherman Fairchild fellowship.
MF is supported by the Zuckerman STEM Leadership Program and in part by the DOE under grant number DE-SC0011640.
ML is supported by the DOE under contract DE-SC0007968 and the Cottrell Scholar Program through the Research Corporation for Science Advancement.
AW is supported by NASA, through ATP grant 80NSSC18K1097 and HST grants GO-14734 and AR-15057 from STScI, and a Hellman Fellowship from UC Davis.
SGK and PFH are supported by an Alfred P. Sloan Research Fellowship, NSF Collaborative Research Grant \#1715847 and CAREER grant \#1455342, and NASA grants NNX15AT06G, JPL 1589742, 17-ATP17-0214.
Numerical simulations were run on the Caltech compute cluster ``Wheeler,'' allocations from XSEDE TG-AST130039 and PRAC NSF.1713353 supported by the NSF, and NASA HEC SMD-16-7592.
This work was performed in part at Aspen Center for Physics, which is supported by National Science Foundation grant PHY-1607611.
We also are grateful for the support from the 2018 CERN-Korea TH Institute.
This research was supported by the Munich Institute for Astro- and Particle Physics (MIAPP) of the DFG cluster of excellence ``Origin and Structure of the Universe".
This research was supported in part by the National Science Foundation under Grant No. NSF PHY-1748958.

RES thanks Nick Carriero, Ian Fisk, and Dylan Simon of the Scientific Computing Core at the Flatiron Institute for their support of the infrastructure housing the the synthetic surveys and simulations used for this work.

This work has made use of data from the European Space Agency (ESA) mission Gaia (\url{http://www.cosmos.esa.int/gaia}), processed by the Gaia Data Processing and Analysis Consortium (DPAC, \url{http://www.cosmos.esa.int/web/gaia/dpac/consortium}). Funding for the DPAC has been provided by national institutions, in particular the institutions participating in the Gaia Multilateral Agreement.

Funding for RAVE has been provided by: the Australian Astronomical Observatory; the Leibniz-Institut fuer Astrophysik Potsdam (AIP); the Australian National University; the Australian Research Council; the French National Research Agency; the German Research Foundation (SPP 1177 and SFB 881); the European Research Council (ERC-StG 240271 Galactica); the Istituto Nazionale di Astrofisica at Padova; The Johns Hopkins University; the National Science Foundation of the USA (AST-0908326); the W. M. Keck foundation; the Macquarie University; the Netherlands Research School for Astronomy; the Natural Sciences and Engineering Research Council of Canada; the Slovenian Research Agency; the Swiss National Science Foundation; the Science \& Technology Facilities Council of the UK; Opticon; Strasbourg Observatory; and the Universities of Groningen, Heidelberg and Sydney.
The RAVE web site is at \url{https://www.rave-survey.org}.

\bibpunct{(}{)}{;}{a}{}{,} 
\def\bibsection{}
\bibliographystyle{aa}
\bibliography{Gaia_ML.bib}

\newpage
\onecolumn

\appendix

\setcounter{equation}{0}
\setcounter{figure}{0}
\setcounter{table}{0}
\setcounter{section}{0}
\makeatletter
\renewcommand{\theequation}{S\thesection.\arabic{equation}}
\renewcommand{\thefigure}{S\thesection.\arabic{figure}}
\renewcommand{\thetable}{S\thesection.\arabic{table}}

\clearpage
\section{\insitu stars.}
\label{app:InSitu}
\begin{figure}[h]
\centering
\includegraphics[width=\linewidth]{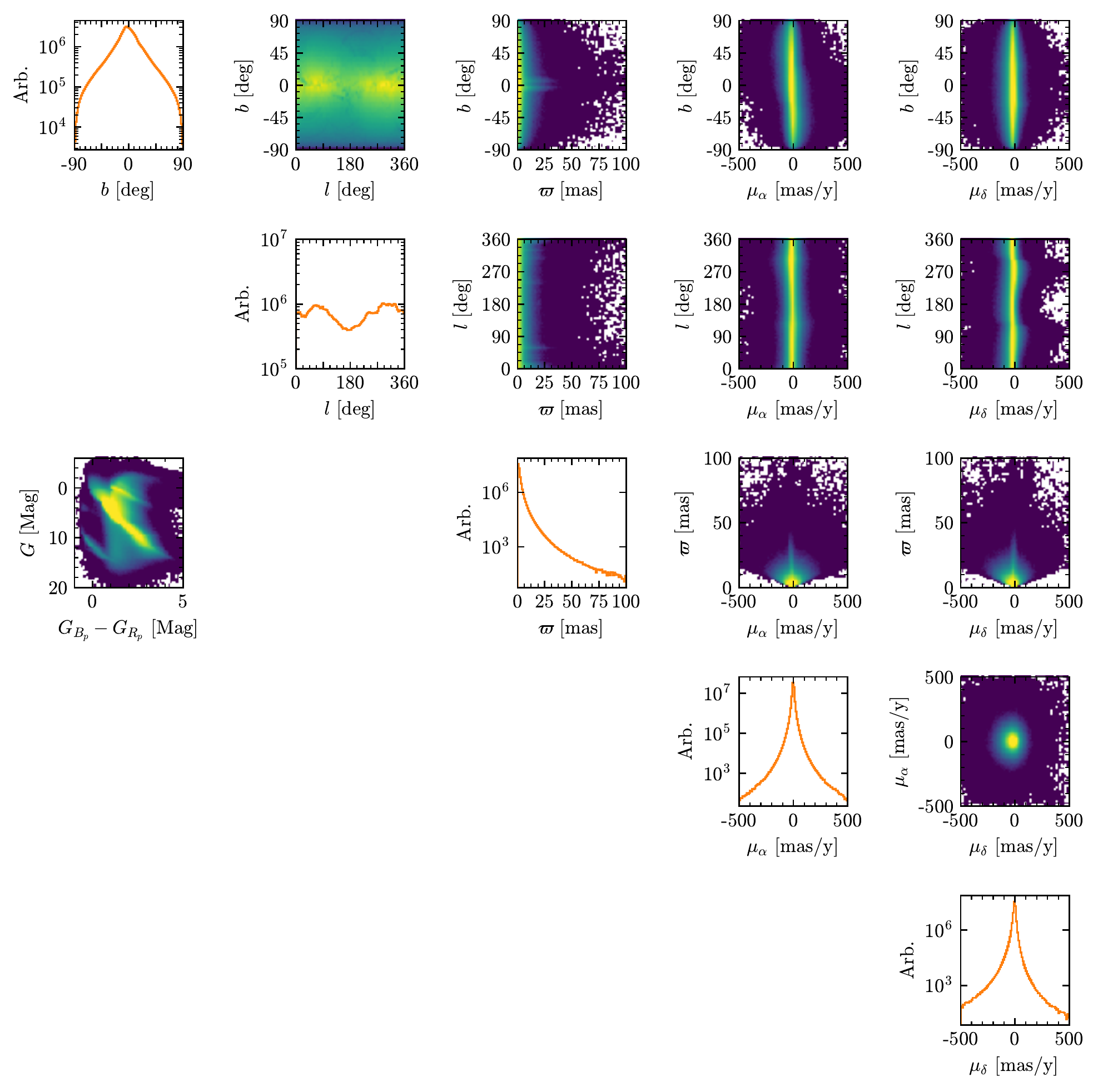}
\caption{Stars within \Gaia DR2 with $\delta\varpi/\varpi < 0.10$.
These stars have been identified as \insitu by our neural network.
The $b$ distribution is strongly peaked at 0 and the $l$ distribution is relatively flat, consistent with stars that make up the disk.
See \cref{Fig:AccretedKinematicAll} in the main text for the analogous distributions of the stars classified as accreted.
}
\label{Fig:InSituKinematicAll}
\end{figure}

The distributions of the stars marked as \insitu by the network are shown in \cref{Fig:InSituKinematicAll}. The disk is easy to see in the $l$--$b$ plane.

\clearpage
\section{Accounting for Measurement Uncertainties}
\label{app:Errors}

\begin{figure}[h]
\includegraphics[width=\linewidth]{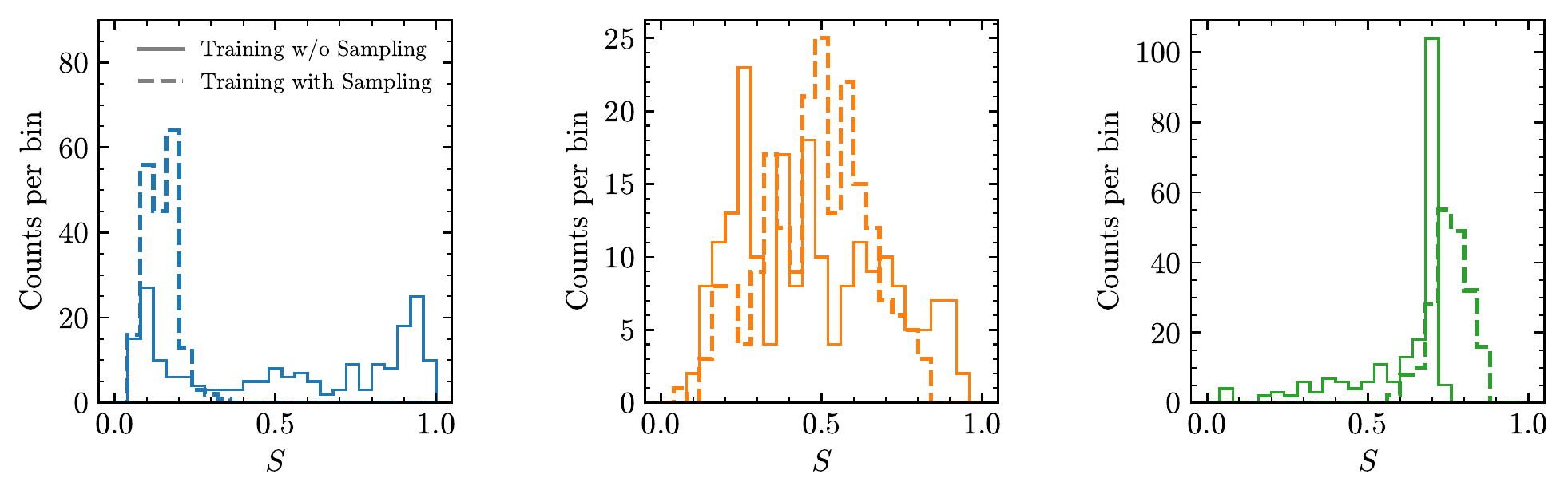}
\caption{Each panel shows a star from the \mi LSR0 test set.
Each of the three panels are computed by taking 200 samples of a single star, where the distribution is derived using the quoted uncertainties in the mock dataset.
The dashed and solid lines show how the network output changes if it was trained with or without error sampling, respectively.
}
\label{Fig:ErrorSampling}
\end{figure}

Given the finite resolution of \Gaia, it is critical to incorporate the measurement uncertainties into our approach.
Fortunately, the \FIRE mock catalogs include uncertainties in the ``measured'' phase space of the simulated stars.
This provides us with the ability to test to what extent it is important to include these uncertainties when training the neural network.
In \cref{Fig:ErrorSampling}, three stars from the test set of \mi LSR0 are shown.
The distributions are made by randomly sampling the stars over their uncertainties and then applying the given network 200 times.
The pairs of distributions shown for each star correspond to the neural network outputs for two different approaches to training.  
The solid lines are computed using the network that was trained without any error sampling in the training step.
In the left and middle panels, the network is very uncertain, and we see there is support across the entire range between 0 and 1.
In the right panel, there is still support over a large range, although most of the random samples fall above a score of 0.5.

The dashed lines show the networks with the configuration used in this work, in which error sampling is included as part of the training step.
Specifically, for each training epoch, a star in the training set is randomly sampled 20 times within its uncertainties.
As the networks typically take around 50 epochs to finish training, the network ultimately sees around 1000 different random samplings per star.
In the left panel, the star that the unsampled network was very uncertain about now has a strong preference for low network scores.
In the middle panel, the network is still not very certain, but the star has lost support at scores as high as 1.
In contrast to this, in the right panel, the star shifts to even larger network outputs after seeing many samples during training.
We interpret this right panel as showing how a truth-level accreted star would be correctly classified by the network that was trained with error sampling, and would have been otherwise missed by the approach that did not include the impact of errors.

We note that the error sampling used here uses uncorrelated uncertainties, as provided in the \FIRE catalogs. 
However, \Gaia DR2 uncertainties are not so diagonal and include correlations. 
Given our tight parallax uncertainty cuts, we do not expect this to be an issue. 
Sampling uncorrelated errors during training should still lead to more precise network scores than not sampling.

\section{Kinematic + Photometric network results}
\label{app:PhotometricResults}
In Sec.~\ref{sec:distance}, we showed that the networks that had access to photometeric information as an input were not able to generalize from bright, well measured stars (\ie, those with 6D phase-space measurements) to dimmer stars without line-of-sight velocity measurements (\ie, those with 5D phase-space measurements).
This implied that only training on real stars for which we could infer empirical labels (implying that they are nearby and bright) would not work for extending to a more general catalog.
However, with transfer learning, we are able to train on simulated stars that which do not have a measurement of the line-of-sight velocity.
This motivates constructing a catalog from a neural network that includes kinematic and photometric inputs.

\begin{figure*}[t]
    \centering
    \includegraphics[width=0.95\textwidth]{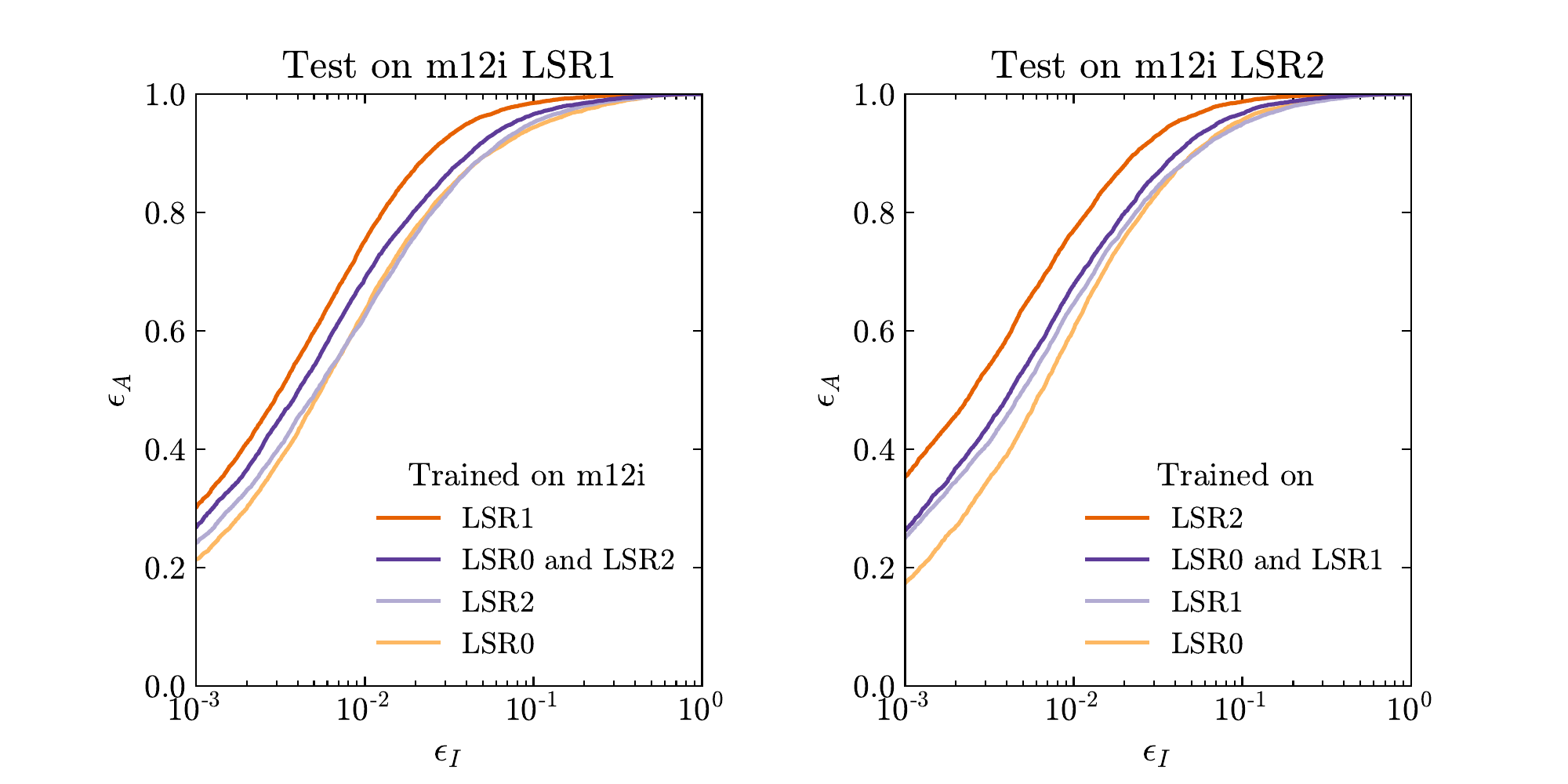}
    \caption{ROC curves showing the performance of networks trained on one LSR of \mi~and applied to a different LSR.
    The networks are trained and tested on the subset of stars in the catalogs that have small parallax errors and use the photometric information in addition to the 5D kinematic information.
    The photometric data generalizes better between catalogs, so the reduction in performance is not as bad as the kinematics-only networks shown in \cref{Fig:MultipleLSRKinematics}.}
    \label{Fig:MultipleLSRKinAndPhot}
\end{figure*}

\Cref{Fig:MultipleLSRKinAndPhot} shows the results of training and testing kinematic + photometric networks on different viewpoints of the \mi galaxy, always using $\delta\varpi / \varpi < 0.10$, such that the photometric distributions should be similar.
Unlike the kinematic networks (\cref{Fig:MultipleLSRKinematics}), the networks trained on multiple LSRs seem to always do better than training on either of the ``wrong'' LSRs alone.
The fractional difference between the orange and purple lines is 8\% and 12\% at a false positive rate of 0.01, which is very similar to that of the kinematic networks.
We conclude that although the photometic networks do not generalize well when extrapolating from brighter to dimmer stars, they do generalize as well as the kinematic-only networks when applied to different viewpoints.

Encouraged by these findings, we repeat the transfer learning methodology tests presented in Sec.~\ref{Sec:Transfer} using 200,000 stars from \mf~LSR0 that satisfy $\delta \varpi / \varpi < 0.10$ and have radial velocity measurements.
The \textbf{ZM selection} is again found to be the best empirical label for transferring the pre-trained networks while only updating the last layer of the network.
The optimal cut value to match the truth-level accreted star velocity distributions is 0.9.
When the method is validated against \mf~LSR1, the results are consistent.

\begin{figure*}[t]
\begin{center}
\includegraphics{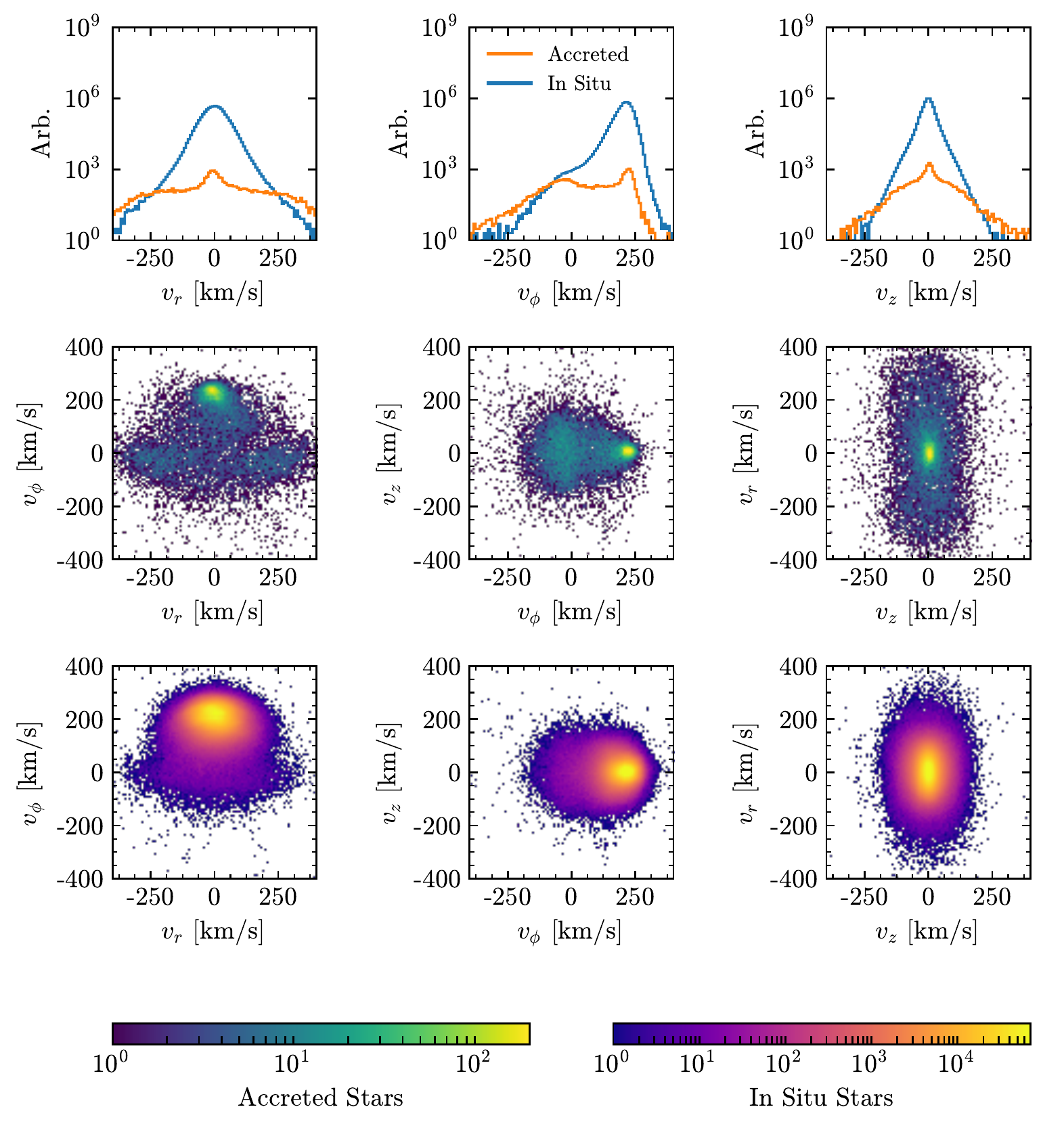}
\caption{Galactocentric velocity distributions of the two  sets of stars in the \Gaia DR2 dataset with small parallax errors and a measurement of the radial velocity as classified by the kinematic + photometric network.   
The  2$^\text{nd}$ and 3$^\text{rd}$ rows provide correlation plots for the stars labeled as accreted and \insitu by the kinematic + photometric network, respectively.  
Many of the selected stars have kinematics similar to the disk.}
\label{Fig:PhotometricRV_AllStars}
\end{center}
\end{figure*}

For the final classifier that will be applied to the real \Gaia data, we start with the kinematic + photometric networks pre-trained on all of the stars from \mi~LSR0, LSR1, and LSR2 with $\delta\varpi / \varpi < 0.10$. 
Transfer learning is performed using the stars of the RAVE~DR5-\Gaia~DR2 cross-matched catalog, using the optimal method discussed above to determine empirical labels.
Overall, the network using photometric information identifies about half as many accreted candidate stars as does the network only using kinematics, 383k compared to 767k.
In addition, the photometric network seems to have more contamination from disk stars than expected.
For example, \Cref{Fig:PhotometricRV_AllStars} shows the subset of stars with 3D velocity measurements.
The most concerning feature is that each of the distributions in the top panels has a peak consistent with the disk dynamics.
In the middle panels, where the 2D velocity distributions of the selected stars are shown, we observe large densities in the same location as the peaks in the \insitu stars.
This is reminiscent of what is observed in \citet{2019A&A...621A..13V}; although their boosted decision tree did not select these specific stars, they did tag a collection of stars whose colors are consistent with metal-poor stars but have kinematics consistent with the disk.
We interpret these peaks as telling us that the kinematic + photometric network is relying too heavily on the photometric information, which could be pointing to an inconsistency between the mock catalogs and the real Milky Way stars tagged as accreted by this network.

\begin{figure*}[t]
\begin{center}
\includegraphics[width=\linewidth]{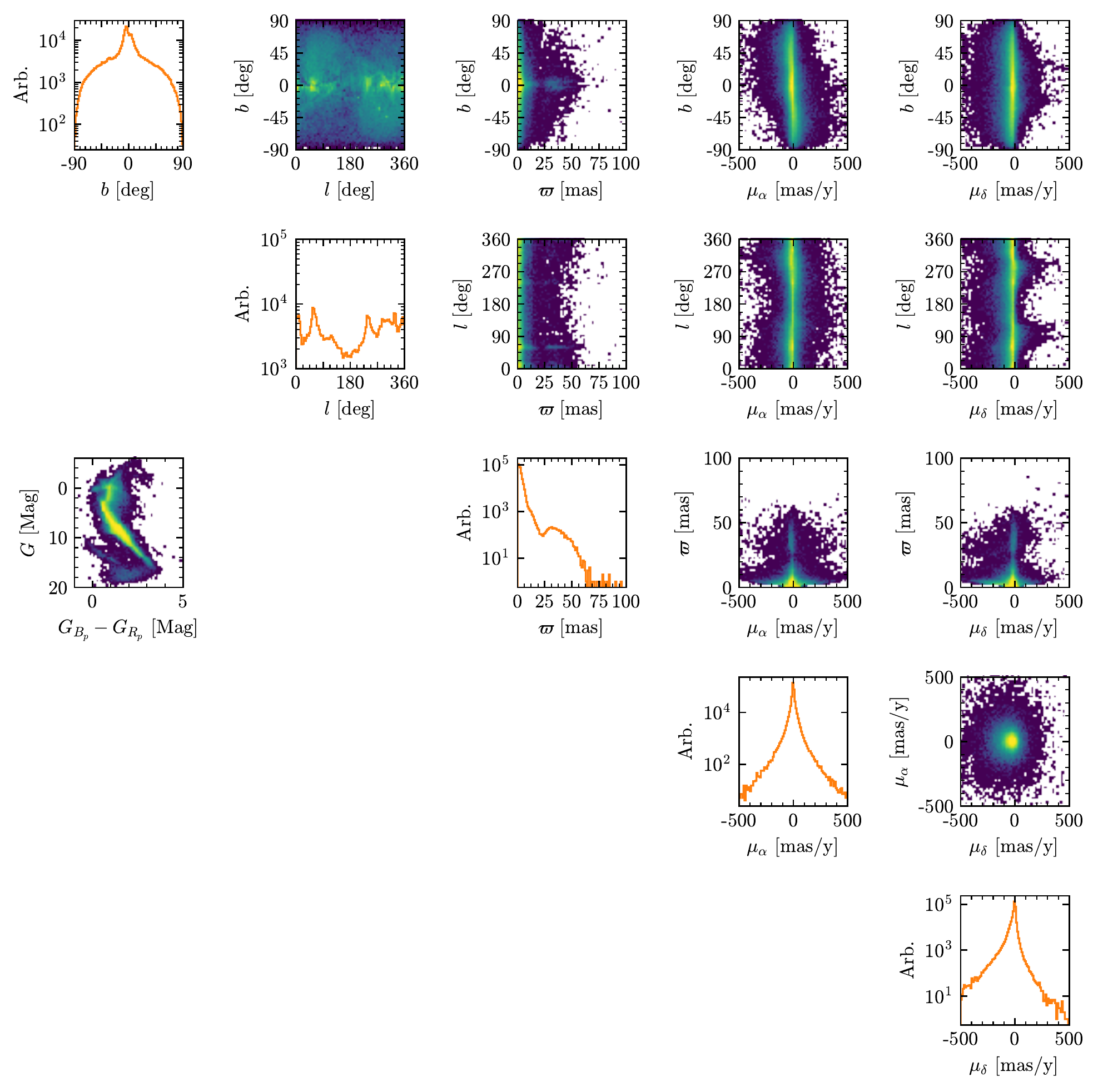}
\caption{Stars from the entire set of stars within \Gaia DR2 with $\delta\varpi/\varpi < 0.10$ that are classified as accreted by the kinematic + photometric network.  See \cref{Fig:AccretedKinematicAll} for the analogous figure for the catalog derived using the kinematic network.}
\label{Fig:AccretedPhotometricAll}
\end{center}
\end{figure*}

Finally, distributions for all of the stars within the \Gaia DR2 dataset that have $\delta \varpi / \varpi < 0.10$ and are classified by the kinematic + photometric network as accreted are shown in \cref{Fig:AccretedPhotometricAll}.
No coordinate transformations have been done, \ie, these are the distributions of the inputs that the network is provided with (other than to the HR diagram).
Comparing with the similar results for the kinematic network shown in \cref{Fig:AccretedKinematicAll}, the kinematic + photometric network selects many more stars at larger parallax and small proper motions. 
The distribution in the HR diagram is also more restricted, and in the $l$--$b$ plane, there is certainly what looks to be a large disk-like component.
We conclude that while there might be interesting features worth exploring within the catalog computed by the kinematic + photometric network, great care should be taken when interpreting any results inferred from it.

\end{document}